\documentclass[aps,
               pra,
               twocolumn,
               showpacs,
               amsmath,
               amssymb,
               superscriptaddress,
               reprint,
               10pt,
              ]{revtex4-2}

\usepackage{physics}
\usepackage{graphicx}  
\usepackage{dcolumn}   
\usepackage{bm,amssymb,amsmath,dsfont}   
\usepackage[colorlinks=true,breaklinks=true,allcolors=blue]{hyperref}
\usepackage{url}
\usepackage{txfonts}
\usepackage{bbold}

\usepackage{graphics}
\usepackage{pstricks}
\usepackage{tikz}
\usepackage{color}
\usepackage{xcolor}
\graphicspath{{Figures/}}

\usepackage{xr}

\renewcommand{\vec}[1]{\mathbf{#1}}
\newcommand{\eq}[1]{(\ref{eq:#1})}
\newcommand{\Eq}[1]{Eq.\,\eqref{eq:#1}}

\newcommand{\Fig}[1]{Fig.~\ref{fig:#1}}
\newcommand{\fig}[1]{\ref{fig:#1}}

\newcommand{\Sect}[1]{Sect.~\ref{sec:#1}}
\newcommand{\sect}[1]{\ref{sec:#1}}

\newcommand{\App}[1]{App.~\ref{app:#1}}

\definecolor{applegreen}{rgb}{0.55, 0.71, 0.0}
\definecolor{byzantine}{rgb}{0.74, 0.2, 0.64}

\newcommand{\cred}[1]{{\color{red}{#1}}}
\newcommand{\cblu}[1]{{\color{blue}{#1}}}

\newcommand{\corg}[1]{{\color{orange}{((TG: #1))}}}
\renewcommand{\corg}[1]{{\color{orange}{}}}

\newcommand{\tbd}[1]{\cred{#1}}
\renewcommand{\tbd}[1]{}

\newcommand{\tbdso}[1]{\cblu{#1}}
\renewcommand{\tbdso}[1]{}

\newcommand{\IntermediateStep}[1]{&\cred{\textrm{\ (($===$ intermediate calc. steps $==>$))}}\nonumber\\ #1  
                                                           &\cred{\textrm{(($<=====================$))}}\nonumber\\}
\renewcommand{\IntermediateStep}[1]{}

\newcommand{\bae}[2]{#1}

\makeatletter
\let\cat@comma@active\@empty
\makeatother

\begin{document}

\title{Non-thermal fixed points of universal sine-Gordon coarsening dynamics}
\author{Philipp Heinen}
\affiliation{Kirchhoff-Institut f\"ur Physik,
             Ruprecht-Karls-Universit\"at Heidelberg,
             Im~Neuenheimer~Feld~227,
             69120~Heidelberg, Germany}
             
\author{Aleksandr N. Mikheev}
\affiliation{Kirchhoff-Institut f\"ur Physik,
             Ruprecht-Karls-Universit\"at Heidelberg,
             Im~Neuenheimer~Feld~227,
             69120~Heidelberg, Germany}
\affiliation{Institut f\"{u}r Theoretische Physik,
		Universit\"{a}t Heidelberg, 
		Philosophenweg 16, 
		69120 Heidelberg, Germany}

\author{Christian-Marcel Schmied}
\affiliation{Kirchhoff-Institut f\"ur Physik,
             Ruprecht-Karls-Universit\"at Heidelberg,
             Im~Neuenheimer~Feld~227,
             69120~Heidelberg, Germany}

\author{Thomas Gasenzer}
\email{t.gasenzer@uni-heidelberg.de}
\affiliation{Kirchhoff-Institut f\"ur Physik,
             Ruprecht-Karls-Universit\"at Heidelberg,
             Im~Neuenheimer~Feld~227,
             69120~Heidelberg, Germany}
\affiliation{Institut f\"{u}r Theoretische Physik,
		Universit\"{a}t Heidelberg, 
		Philosophenweg 16, 
		69120 Heidelberg, Germany}

\date{\today}

\begin{abstract}
We examine coarsening of field-excitation patterns of the sine-Gordon (SG) model, in two and three spatial dimensions, identifying it as universal dynamics near non-thermal fixed points.
The SG model is relevant in many different contexts, from solitons in quantum fluids to structure formation in the universe.
The coarsening process entails anomalously slow self-similar transport of the spectral distribution of excitations towards low energies, induced by the collisional interactions between the field modes.
The focus is set on the non-relativistic limit exhibiting particle excitations only, governed by a Schr\"odinger-type equation with Bessel-function non-linearity.
The results of our classical statistical simulations suggest that, in contrast to wave turbulent cascades, in which the transport is local in momentum space, the coarsening is dominated by rather non-local processes corresponding to a spatial containment in \emph{position} space.
The scaling analysis of a kinetic equation obtained with path-integral techniques corroborates this numerical observation and suggests that the non-locality is directly related to the slowness of the scaling in space and time.
Our methods, which we expect to be applicable to more general types of models, could open a long-sought path to analytically describing universality classes behind domain coarsening and phase-ordering kinetics from first principles, which are usually modelled in a near-equilibrium setting by a phenomenological diffusion-type equation in combination with conservation laws.
\end{abstract}

\maketitle

\emph{Introduction}. 
Phase-ordering kinetics and coarsening following a quench into a phase with different order have been studied since a  long time \cite{Lifshitz1961a,Lifshitz1962a,Wagner1961a,
Bray1994a.AdvPhys.43.357,Puri2019a.KineticsOfPT,Cugliandolo2014arXiv1412.0855C}.
The standard classification of such phenomena is closely related to that of dynamical critical scaling in the linear response of systems out of but close to equilibrium \cite{Hohenberg1977a,Bray2003a.rsta.2002.1164}, as well as to non-linear critical relaxation \cite{Racz1975a.PLA.53.433,Fisher1976a.PhysRevB.13.5039,Bausch1976b,Bausch1979a}.
Coarsening, though, in general represents a dynamical process, which results from a quench far out of equilibrium and exhibits critically slowed evolution.
For example, the average size of spin domains forming in a shock-cooled magnet, on long time scales, grows as a power law in time, $\ell_{\mathrm{d}}(t)\sim t^{\,\beta}$, with a universal scaling exponent $\beta$.
Moreover, the momentum-space distribution of the order parameter field is typically character\bae{ise}{ize}d by a universal, so-called Porod tail, $f(t,p)\sim p^{-\kappa}$. 
Here, universality means that the spatio-temporal form of this distribution and of more general statistical correlations is independent of the particular physical realisation and reflects characteristic symmetries and related conservation laws. 

Coarsening is typically described by diffusion-type equations for the order parameter distribution, such as the Allen-Cahn or the Cahn-Hilliard equations, for non-conserved and conserved order parameters, respectively, which yield a relative scaling in space and time as set by the exponent $\beta$  \cite{Bray1994a.AdvPhys.43.357}.
Beyond this diffusive picture, the growth law, together with the universal shape of the distribution and associated conservation laws imply that coarsening involves a non-linear transport process in momentum space towards lower $p$.
In this respect it is similar to the build-up of inverse cascades in wave-turbulence \cite{Zakharov1992a,Nazarenko2011a}, as well as in classical \cite{Kraichnan1967a,Frisch1995a} and superfluid turbulence \cite{Vinen2006a,Tsubota2008a}.
During recent years, studies of universal phenomena far from equilibrium have intensified, in experiment
\cite{%
Navon2015a.Science.347.167N,
Navon2016a.Nature.539.72,
Eigen2018a.arXiv180509802E,
Prufer:2018hto,
Erne:2018gmz,
Navon2018a.doi:10.1126/science.aau6103,
Glidden:2020qmu,
GarciaOrozco2021a.PhysRevA.106.023314}
and theory
\cite{%
Berges:2008wm,
Schole:2012kt,
Nowak:2012gd,
Hofmann2014a,
Maraga2015a.PhysRevE.92.042151,
Orioli:2015dxa,
Williamson2016a.PhysRevLett.116.025301,
Williamson2016a.PhysRevA.94.023608,
Bourges2016a.arXiv161108922B.PhysRevA.95.023616,
Chiocchetta:2016waa.PhysRevB.94.174301,
Karl2017b.NJP19.093014,
Schachner:2016frd,
Walz:2017ffj.PhysRevD.97.116011,
Mikheev:2018adp,
Schmied:2018upn.PhysRevLett.122.170404,
Mazeliauskas:2018yef,
Schmied:2018mte,
Williamson2019a.ScPP7.29,
Schmied:2019abm,
Gao2020a.PhysRevLett.124.040403,
Wheeler2021a.EPL135.30004,
Gresista:2021qqa,
RodriguezNieva2021a.arXiv210600023R,
Preis:2022uqs}, 
many of them in the field of ultracold gases.
In their light, a rigorous renormal\bae{isation}{ization}-group (RG) analysis, including a comprehensive classification scheme of non-linear, far-from-equilibrium universal dynamics would be of strong interest, but is lacking so far for most practically relevant cases of coarsening \cite{Cugliandolo2014arXiv1412.0855C}.
It has been proposed that, in analogy to the characterisation of equilibrium critical phenomena, non-thermal fixed points \cite{Berges:2008wm} exist that account for universal scaling dynamics far from equilibrium 
\cite{%
Orioli:2015dxa,
Prufer:2018hto,
Erne:2018gmz,
Schmied:2018mte,
RodriguezNieva2021a.arXiv210600023R,
Preis:2022uqs}, 
including phase-ordering and coarsening \cite{Schole:2012kt,Nowak:2012gd,Karl2017b.NJP19.093014}. 

\begin{figure}[t]
	\includegraphics[width=0.8\columnwidth]{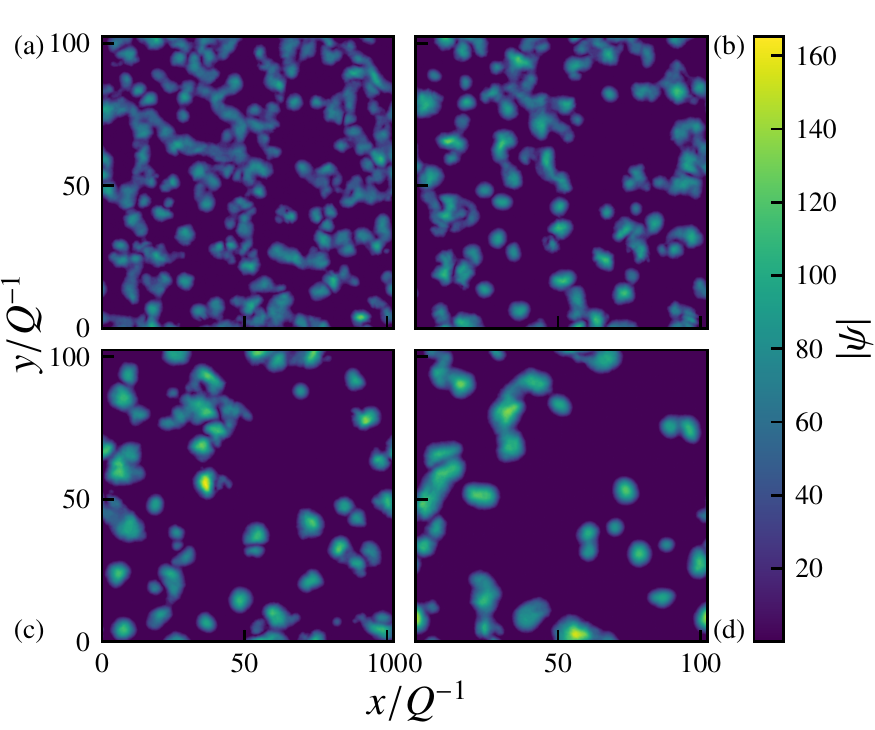}
	\includegraphics[width=0.8\columnwidth]{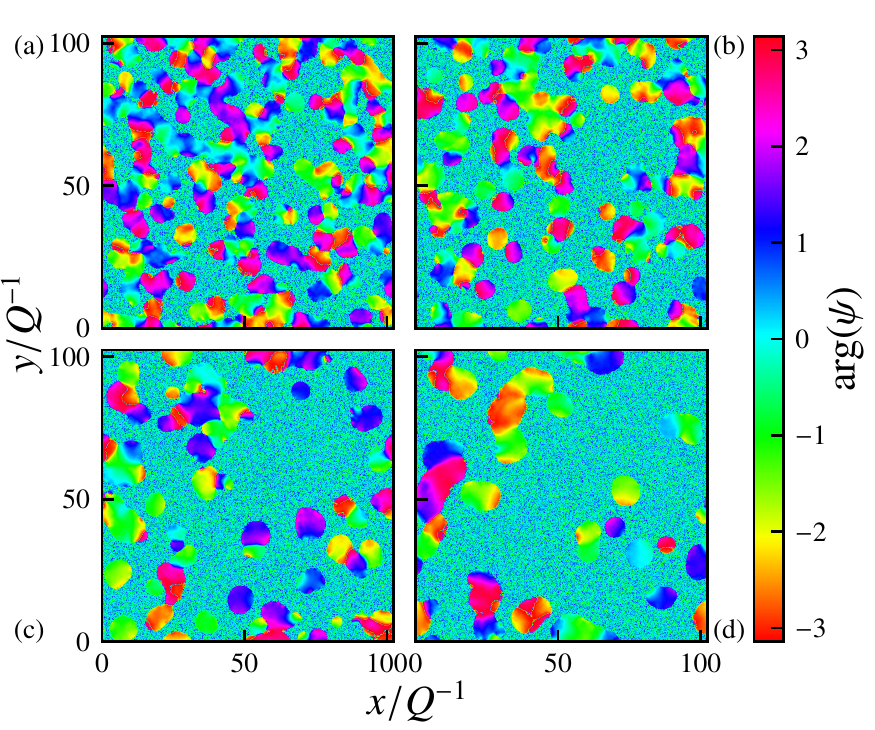}
	\caption{Amplitude $|\psi(\mathbf{x},t)|$ (upper) and phase angle $\arg[\psi(\mathbf{x},t)]$ (lower panels) of the time evolving field distribution in position space, $\mathbf{x}=(x,y)$, for a single run of the simulations in $d=2$ dimensions, for $m/Q=20$, $F_{0}=10^{3}$, at  times (a) $10^{-3}Q\,t=0.5$, (b) $1.0$, (c) $2.0$, and (d) $4.0$. 
	Whereas the amplitude remains $|\psi|\simeq0$ in many places, it grows large inside small separated patches within which the phase angle $\mathrm{arg}(\psi)$ remains approximately constant.
	During the evolution, the phase varies between patches, which continuously merge and grow, giving rise to universal coarsening dynamics, which manifests itself as rescaling in space and time \cite{Videos}, cf.~\Fig{OccupationNumber2D}.
	}
	\label{fig:RealSpace2D}
\end{figure}

\emph{Main result.} 
We study numerically, by means of statistical simulations, coarsening dynamics of field-excitations in the non-relativistic limit of the sine-Gordon (SG) model and compare it with analytically predicted scaling exponents.
The SG model is relevant in many contexts, including soliton and kink solutions, its mapping to a Coulomb gas, e.g., in describing the Berezinskii-Kosterlitz-Thouless transition \cite{Gogolin2004a.bosonization,Giamarchi2003a,Cuevas2014a.sineGordon,Minnhagen1987a.RevModPhys.59.1001,Nandori2022a.NPB975.115681}, as well as 
 structure formation and growth in the universe \cite{Turok1991a.PhysScr,Greene:1998pb,Berges:2014xea,Berges:2019dgr,Lentz2019a.MNRAS485.1809}
 and false vacuum decay \cite{Hawking1982a.PhysRevD.26.2681,Braden2015a.JCAP03.007}.
   
Our simulations start from initial states dominated by fundamental particles while antiparticles are neglected.
We extract the exponents $\alpha$, $\beta$, and $\kappa$ governing the universal scaling form and evolution of the structure factor, i.e.~the order-parameter spectrum with respect to a reference time $t_{0}$,
\begin{align}
	\label{eq:scalrel}
	f(t,\mathbf{p})
	&=(t/t_0)^\alpha f_\mathrm{s}\left([t/t_0]^\beta \mathbf{p}\right)
	\,,\\
	\label{eq:porodlaw}
	f_\mathrm{s}(\mathbf{p})
	&\sim |\mathbf{p}|^{-\kappa}\,,
	\qquad |\vec p|\gg p_{\Lambda}\sim (t/t_{0})^{-\beta}
	\,.
\end{align}
where $p_{\Lambda}(t)$ is a dynamically generated infrared (IR) cutoff scale. 
We compare these with recent analytical predictions \cite{Heinen2022b} for coarsening according to the SG non-linear dynamic equation for the scalar order-parameter field $\varphi$,
\begin{equation}
	\label{eq:sGE}
	\Box\varphi(x) + m^{2} \sin \varphi(x) = 0
	\,, 	
\end{equation}
where $x=(t,\vec x)$. 
In $d$ spatial dimensions, they are found to be
\begin{align}
	\label{eq:exponents}
	\alpha=d\beta
	\,,\qquad
	\beta=\frac{1}{2+d}
	\,,\qquad
	\kappa=2d+2
	\,.
\end{align}
The exponent $\alpha$ accounts for the dynamical scaling dimension of the function $f_\mathrm{s}$, and $\alpha=d\beta$ implies the momentum integral over $f$ to be conserved in time. 

Our numerical results demonstrate that the power-law fall off \eq{porodlaw} prevails within a finite region of momenta $p\equiv|\vec p|$ only, with $0\leq p\lesssim p_{\lambda}$, where $p_{\lambda}\gg p_{\Lambda}$ marks a crossover to a nearly thermal\bae{ise}{ize}d tail. 
Hence, the scaling function for $p\lesssim p_{\lambda}$, is found to be well approximated by the typical form
\begin{align}
	\label{eq:fMK}
	f_\mathrm{s}(\mathbf{p})\sim \frac{p_{\Lambda}^{\kappa-d}}{p_{\Lambda}^{\kappa}+p^{\,\kappa}}
	\,,
\end{align}
in which the IR cutoff rescales according to $p_{\Lambda}(t)\sim t^{-\beta}$, such that $f(t,\vec p)$ satisfies \eq{scalrel}.
Note that, as is always implied, the high-momentum tail, which takes a thermal form, does not contribute to the scaling but is understood to act as an energy sink.
We emphasise that the exponents \eq{exponents} are substantially smaller (larger) than $\alpha/d=\beta=1/2$ ($\kappa=d+1$) at the `Gaussian' \cite{Mikheev:2018adp} non-thermal fixed point dominated by the universal redistribution of phase excitations \cite{Orioli:2015dxa,Chantesana:2018qsb.PhysRevA.99.043620,Mikheev:2018adp}.

\emph{Simulations.}
We study universal dynamics by means of numerical simulations at low energies, where the SG dynamics can be described by a non-linear Schr\"odinger model with Bessel-function non-linearity \cite{Eby:2014fya,Braaten:2016kzc,Robson2021a.arXiv210504498R} with Hamiltonian
\begin{align}
	\label{eq:GPBHamiltonian}
	{H}_{\text{GPB}}
	=&\int \mathrm{d}^{d}\!x\left\{
	-\frac{\psi^*\Delta\psi}{2m}
	- {2m}\left[J_0(|\psi|)+\frac{1}{4}|\psi|^2-1\right]\right\}
	\,,
\end{align}
cf.~App.~\sect{NRLimitSG}, 
giving rise to the non-linear equation of motion
\begin{align}
	\label{eq:GPBE}
	\mathrm{i}\partial_{t}\psi
	=-\frac{1}{2m}\Delta \psi +m\left(\frac{1}{|\psi|}J_1(|\psi|)-\frac{1}{2}\right)\psi
	\,,
\end{align}
for a complex field $\psi\in\mathbb{C}$ defined by $\varphi=\Re\{\psi\exp(-\mathrm{i}mt)\}$.
For the system to approach a non-thermal fixed point, we initial\bae{ise}{ize} it far from equilibrium.
To that end we choose the initial momentum distribution $\langle \lvert \psi(\mathbf{p},0) \rvert^2\rangle\equiv f(t=0,\mathbf{p}=0)\,\Theta(Q-p)$ to be constant up to a cutoff $Q$.
We compute correlations within a Truncated-Wigner (TW) approach, adding, in each run, Gaussian noise of half a particle per mode to the initial distribution and choosing the phases $\theta(\mathbf{p},0)$ of the complex field $\psi=f^{1/2}\exp(\mathrm{i}\theta)$ randomly on the circle \cite{Blakie2008a,Polkovnikov2010a}.

\begin{figure}
	\centering
	\includegraphics[width=0.95\columnwidth]{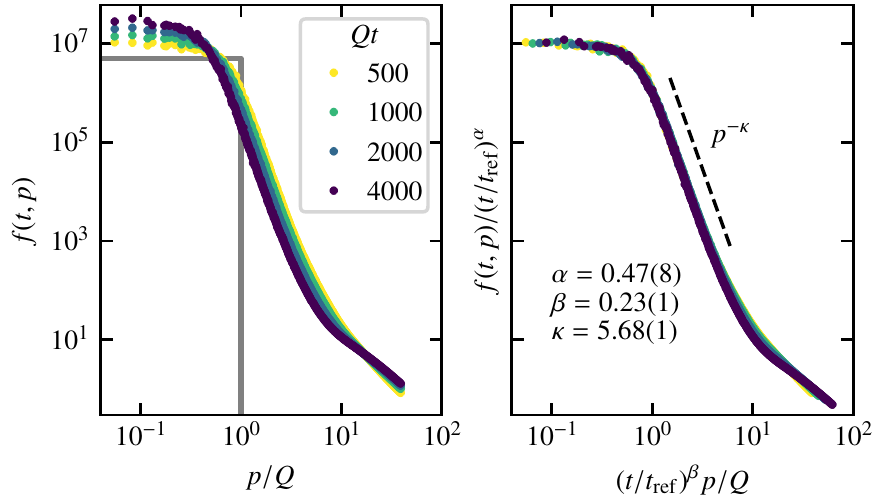}
	\caption{(Left panel) Late-time evolution of $f(t,p)$ obtained by averaging over $25$ runs and angular orientations $\mathbf{p}/p$, for the same time steps as depicted in \Fig{RealSpace2D}, from the initial distribution (grey solid line). 
	(Right panel) Self-similar scaling collapse  according to \eq{scalrel}, with exponents $\alpha=0.47(8)$ and $\beta=0.23(1)$, with respect to $t_{\mathrm{ref}}=500\,Q^{-1}$.
	In the power-law region, the spectrum falls off as $f_{\mathrm{s}}(p)\sim p^{-\kappa}$, with $\kappa=5.68(1)$. 
	}
	\label{fig:OccupationNumber2D}
\end{figure}

\Fig{RealSpace2D} shows four snapshots of the late-time evolution in a single exemplary run in $d=2$, with periodic boundary conditions, depicting the spatial pattern of the amplitude $|\psi|$ as well as of $\theta$.
The strength of the excitations is quantified by the dimensionless ratio $F_{0}\equiv\eta n_{0}/M$ in terms of the quasiparticle density $n_{0}=\int_{{\vec q}_{i}}f_{{\vec q}_{i}}$ and a parameter $\eta$ that is introduced because the field $\psi$ is dimensionless in position space, see \App{SGmodel}.
For videos of single runs  \footnote{In the videos, a fast rotating phase has been divided out by subtracting the term $-m\,\psi/2$ from \eq{GPBE}.} in $d=2$ and $3$ spatial dimensions cf.~\cite{Videos}.
Corresponding results in $d=3$ are summar\bae{ise}{ize}d in Appendix \sect{NumResults}.

The panels show that the amplitude $|\psi|$, within an increasing part of the volume, remains trapped in the zeroth minimum of the Bessel-function field potential, $\psi\simeq0$, with highly fluctuating phase, but grows large within separated patches. 
While the phase angle is approximately constant within one patch, it varies randomly between the patches. 
The flatness of the phase within a patch is due to the amplitude taking large values there, such that the contribution from the Bessel function to the potential energy in \eq{GPBHamiltonian} approximately vanishes.
Hence, the shape of the field $\psi$ within a large part of the patch is dominated by the free Schr\"odinger equation with a zero-momentum gap $\mu=m/2$.
As a result, the amplitude $|\psi|$ falls off, within a single patch, approximately as a free wave, with wave length on the order of the twice the size $a$ of the patch, which is initially $a\simeq2\pi/Q$.
Moreover, the patches are being deformed during mutual mergers, and their average size grows, giving rise to a self-similar coarsening evolution of the system which manifests itself as rescaling in space and time.

\Fig{OccupationNumber2D} depicts the time evolution of $f(t,{\vec p})$ resulting from an angular average over $\mathbf{p}$ as well as $25$ TW runs. 
While the left panel shows the distribution at the same time steps as chosen in \Fig{RealSpace2D},
the right panel demonstrates the rescaling collapse of the distributions, for a fixed set of exponents $\alpha$ and $\beta$.
It thus demonstrates that the evolution follows a universal rescaling according to \eq{scalrel}, \eq{porodlaw}, with a Porod exponent $\kappa=5.68(1)$.
See App.~\sect{NumResults} for more details.
The values for the scaling exponents are remarkably close to our analytical predictions \eq{exponents}.

\begin{figure}
	\centering
	\includegraphics[width=0.95\columnwidth]{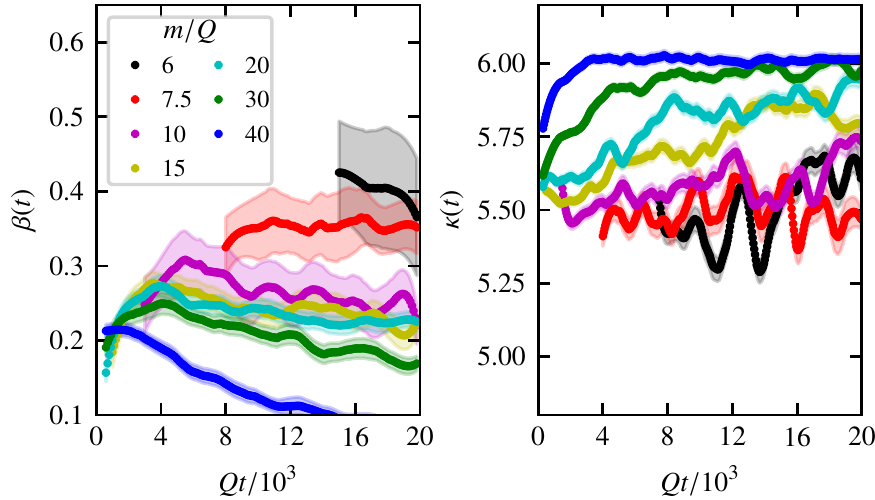}
	\caption{(Left panel) Scaling exponent $\beta$, for $d=2$, as a function of time $t$, obtained by performing a scaling collapse of $f(t,p)$ at times $t/2$, $3t/4$, $t$. 
	The resulting evolution (average over $50$ runs)  is shown for $7$ different values of the gap parameter $m/Q$, as given in the legend, smoothed by means of a Savitzky-Golay filter, cf.~App.~\sect{NumResults}. 
	Fluctuations are indicated by shaded color bands.
	(Right panel) The same for $\kappa(t)$, obtained by fitting the form \eq{fMK} to $f(t,p)$, within a range of momenta from the lowest $p$ to one half-way between the two bending scales.
	Both panels demonstrate scaling with approximately constant universal exponents within the range $m/Q\sim15\dots30$.  
	}
	\label{fig:betakappaoft2D}
\end{figure}

Note that, during the early-time evolution, \emph{prescaling} can prevail \cite{Schmied:2018upn.PhysRevLett.122.170404} during which the scaling function has not yet assumed its universal form and the exponents could still change \cite{Mazeliauskas:2018yef}.
In order to obtain a more refined picture of the coarsening dynamics observed here, we performed a time-resolved scaling analysis of the momentum distribution $f(t,p)$.
Specifically, we rescaled, at a given moment $t$, the numerically obtained distributions at times $t/2$, $3t/4$, $t$ onto each other, by choosing $\alpha=d\beta$ and varying $\beta$, thereby minimizing the mean squared difference of the distributions where their arguments overlap, up to a momentum logarithmically half-way between $p_{\Lambda}$ and $p_{\lambda}$ of the distribution at time $t$.
At each time $t$, we fitted a power law \eq{porodlaw} to the averaged distribution between the same momenta, to extract $\kappa$.
In this way, we obtained time-evolving scaling exponents $\beta(t)$, $\kappa(t)$, which are shown in \Fig{betakappaoft2D} for different values of the gap parameter $m/Q$. 
Our results indicate that universal scaling evolution, with approximately constant exponent consistent with $\beta\sim 0.25$ can be observed within a range of $m/Q\simeq15\dots30$. 
At later times, here already seen for the case of $m/Q=40$, the coarsening is observed to come to a halt, after well-separated, randomly distributed, nearly circular patches have formed, see \cite{Videos} for a visuali\bae{s}{z}ation.
At smaller $m/Q$, the exponent $\beta$ increases.
The exponent $\kappa$ is seen to approach its final value the earlier the larger $m/Q$.

We have repeated the same analysis in $d=3$ dimensions, as summar\bae{ize}{ise}d in App.~\sect{NumResults}, for which we found $\alpha=0.85(7)$, $\beta=0.27(1)$, and $\kappa=7.74(1)$. 
These deviate from the analytical predictions \eq{exponents}.
Nevertheless, the exponents, $\beta$ and $\kappa$ are substantially smaller and larger, respectively, than in the Gaussian case, where $\beta=1/2$  and $\kappa=d+1$, cf.~\cite{Orioli:2015dxa,Chantesana:2018qsb.PhysRevA.99.043620,Mikheev:2018adp} and remarkably close to the predictions \eq{exponents}.

\begin{figure}
	\centering
	\includegraphics[width=0.85\columnwidth]{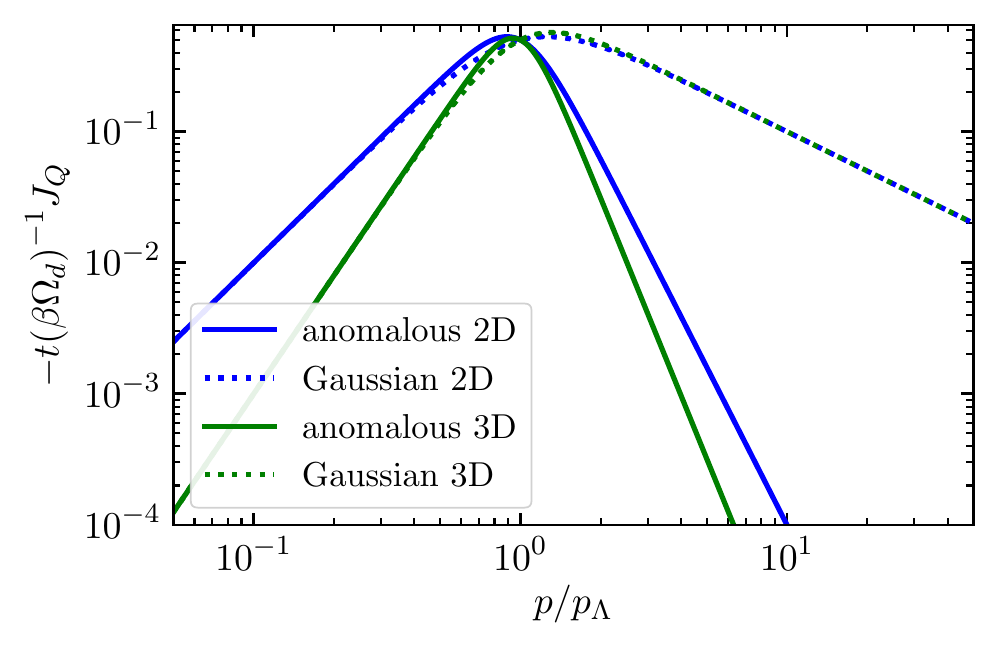}
	\caption{Non-locality of the self-similar transport near a non-thermal fixed point. The figure shows the momentum dependence of the temporally rescaled radial current, $t(d\beta\Omega_{d})^{-1} J_{Q}$, for the anomalous fixed point of the SG model (exponents $\kappa=2d+2$), as compared to the Gaussian one ($\kappa=d+1$), in $d=2$ (solid lines) and $d=3$ (dashed) dimensions.
	The anomalous transport is much more strongly peaked near the IR scale $p_{\Lambda}$ than in the Gaussian case.
	}
	\label{fig:SourcesJQ}
\end{figure}

\emph{Non-locality of transport.}
In contrast to the local, near-forward scattering in wave turbulence independent of the physics at its ultraviolet (UV) and infrared (IR) ends \cite{Zakharov1992a,Nazarenko2011a,Balk1988a,Balk1990a,Balk1990b}, the universal transport here  involves \emph{non-local} scattering in momentum space.
Describing, as standard in wave turbulence, the transport by a continuity equation in momentum space, $0=\partial_{t}N_\mathrm{Q}(t,p)+\partial_{{p}}{J}_\mathrm{Q}(t,p)$ for the radial quasiparticle number $N_\mathrm{Q}(t,p)=\Omega_{d}p^{d-1}f(t,p)$ and radial current ${J}_\mathrm{Q}$ \cite{Zakharov1992a}, we can estimate the locality of the current from its scaling, ${J}_\mathrm{Q}(t,p)\sim (p/t)N_{\mathrm{Q}}(t,p)$.
Inserting the scaling form \eq{fMK} into $N_{\mathrm{Q}}$, one finds that ${J}_\mathrm{Q}(t,p)\sim t^{-2}p^{-d-2}\sim t^{-1} (p/p_{\Lambda})^{-2-d}$ for $p\gg p_{\Lambda}$ while ${J}_\mathrm{Q}(t,p)\sim t^{-2/(d+2)}p^{d}\sim t^{-1} (p/p_{\Lambda})^{d}$ for $p\ll p_{\Lambda}$.
So the current takes the same form ${J}_\mathrm{Q}(t,p)\sim t^{-1} (p/p_{\Lambda})^{\iota}$, with $\iota>0$ ($<0$) below (above) the scale $p_{\Lambda}$.
Hence, in contrast to local cascades of stationary wave-turbulent flows, the radial current is peaked around the characteristic scale $p\approx p_{\Lambda}$ and decreases in time.
Like the distribution \eq{porodlaw}, $J_{\mathrm{Q}}$ depends strongly on the location of the IR cutoff scale $p_{\Lambda}$. 
\Fig{SourcesJQ} illustrates the momentum distribution of the temporally rescaled current $t(d\beta\Omega_{d})^{-1} J_{Q}$, see \Eq{RadialCurrentfromMKf} in the appendix for further details.
The modes just below $p_{\Lambda}$ undergo the strongest growth while all modes $p_{\Lambda}\lesssim p\lesssim p_{\lambda}$ get depleted, the more the closer they are to $p_{\Lambda}$.
In fact, all modes $p\lesssim p_{\Lambda}$ grow, but due to the volume factor, most of the excitations are deposited closer to $p_{\Lambda}$.
For comparison, we also depict the current for the case of the  `Gaussian' non-thermal fixed point \cite{Orioli:2015dxa,Walz:2017ffj.PhysRevD.97.116011,Chantesana:2018qsb.PhysRevA.99.043620} which is also character\bae{ise}{ize}d by a non-local current, though distinctly less concentrated near $p_{\Lambda}$, with a weaker fall-off ${J}_\mathrm{Q}(t,p)\sim t^{-1/2}p^{-1}\sim t^{-1}(p/p_{\Lambda})^{-1}$ for $p\gg p_{\Lambda}$, see \Fig{SourcesJQ}.

\emph{Non-local kinetics.}
The observed non-locality of transport corroborates the analytical description, which lead to the exponents \eq{exponents}   \cite{Heinen2022b}. 
This analysis is based on a non-perturbative kinetic equation governing $f(t,\mathbf{p})$,
\begin{equation}
	\label{eq:KinEq}
	\partial_t f(t,\mathbf{p}) = C[f](t,\mathbf{p})\,.
\end{equation}
The scattering integral $C[f]$ takes a wave-Boltzmann form,
\begin{align}
	\label{eq:Cfmain2}
	&C[f](t,\mathbf{p})
	=-\sum_{n=1}^{\infty}
	\int\prod_{i=1}^{2n+1}\frac{\dd{\mathbf{q}_{i}}}{(2\pi)^{d}} \,
	\left|T^{(n)}(t;{\mathbf{p},\mathbf{q}_{1},\dots,\mathbf{q}_{2n+1}})\right|^{2}
	\nonumber\\
	&\quad\times\
	\delta(\omega_{\mathbf{p}}-\omega_{\mathbf{q}_{1}}-\dots-\omega_{\mathbf{q}_{n+1}}
	+\omega_{\mathbf{q}_{n+2}}+\dots+\omega_{\mathbf{q}_{2n+1}})
	\nonumber\\
	&\quad\times\
	\delta(\mathbf{p}-\mathbf{q}_{1}-\dots-\mathbf{q}_{n+1}+\mathbf{q}_{n+2}+\dots+\mathbf{q}_{2n+1})
	\nonumber\\
	&\quad\times\
	\left[(f_{\mathbf{q}_{1}}+1)\cdots(f_{\mathbf{q}_{n+1}}+1)f_{\mathbf{q}_{n+2}}\cdots f_{\mathbf{q}_{2n+1}}f_{\mathbf{p}}
	\right.
	\nonumber\\
	&\quad\quad-\
	\left.
	f_{\mathbf{q}_{1}}\cdots f_{\mathbf{q}_{n+1}}(f_{\mathbf{q}_{n+2}}+1)\cdots 
	(f_{\mathbf{q}_{2n+1}}+1)(f_{\mathbf{p}}+1)\right]
	\,,
\end{align}
which, following from the Taylor expansion of the cosine potential, includes elastic collisions of modes $\{\mathbf{p},\mathbf{q}_{i}\}$ with frequencies, in the low-energy limit, $\omega_{\vec q}\simeq M+q^{2}/2M$.
In all scattering terms in \eq{Cfmain2} the large rest masses $M$ cancel in the energy conservation condition.
The transfer or $T$-matrix, cf.~App.~\sect{ScattInt},
\begin{align}
	\label{eq:Tmatrix}
	&\left|T^{(n)}(t;\mathbf{p},\{\mathbf{q}_{i}\})\right|^{2}
	= 
	\frac{g_{\mathrm{eff}}^{2}(n;t;\mathbf{p},\{\mathbf{q}_{i}\})}{n!(n+1)!}
	\frac{\eta^{2n+2}}{2\omega_{\mathbf{p}}}
	\prod_{i=1}^{2n+1}\frac{1}{2\omega_{\mathbf{q}_{i}}} 
	\,,
\end{align}
includes, besides an effective, momentum-dependent coupling function $g_{\mathrm{eff}}$, approximately constant factors, $\omega_{\vec p}\simeq M$.
$g_{\mathrm{eff}}$ is found to depend on the distribution $f_\mathbf{p}$ itself and thus to
modify the scaling of the scattering integral in dependence of $\beta$ and $\kappa$.
Proposing the scaling form \eq{scalrel}, \eq{fMK} to solve \eq{KinEq}, one infers the scaling exponents $\alpha$ and $\beta$, as well as, for a fixed time $t_{0}$, $\kappa$, cf.~Refs.~\cite{Chantesana:2018qsb.PhysRevA.99.043620,Heinen2022b} and App.~\sect{ScalingAnalysis}.
This is done analogously as in studying wave turbulence \cite{Zakharov1992a,Nazarenko2011a}.

In contrast to standard cases of wave turbulence \cite{Zakharov1992a,Nazarenko2011a} and non-thermal fixed points \cite{Orioli:2015dxa,Walz:2017ffj.PhysRevD.97.116011,Chantesana:2018qsb.PhysRevA.99.043620}, the above kinetic equation involves elastic scattering of in principle arbitrarily many momentum modes.
It is, in particular, found, that for values of the dimensionless ratio $F_{0}\equiv\eta n_{0}/M$ that are much greater than $1$, the scattering integral \eq{Cfmain2} is dominated by terms of order $n\approx F_{0}$ involving a large number of modes \cite{Heinen2022b}.
While the elastic scattering processes conserve the total energy and momentum of the scattering partners, the particular momentum of a single one amongst the many modes involved remains rather unconstrained. 
Hence, and in contrast to elastic two-to-two scattering (resulting for $F_{0}\approx1$), there is more freedom to fulfil the conservation laws, and more scattering partners have  a momentum on the order of $p_{\Lambda}$.
In fact, up to a small number (3, cf.~\cite{Heinen2022b}), most of the momenta must be $q_{i}\lesssim p_{\Lambda}$. Otherwise, the scattering integral, being a sum of many different $n$, is not a homogeneous function of momentum.
This results in the modified exponents \eq{exponents} as compared to the Gaussian case, and in the non-locality of transport in momentum space illustrated in \Fig{SourcesJQ}, cf.~also App.~\sect{nonlocality-in-kinetics}.

\emph{Summary}.
Coarsening of local\bae{ise}{ize}d field-excitation patterns of the sine-Gordon model in two and three dimensions is found to be character\bae{ise}{ize}d by anomalously slow scaling in space and time.
Remarkably, in contrast to wave-turbulent cascade-like local transport between momentum scales, this self-similar transport is dominated by strongly non-local scattering processes in momentum space, corresponding to a spatial containment in \emph{position} space.
Recent scaling analysis of a kinetic equation obtained with path-integral techniques corroborates this numerical observation and suggests that the non-locality is directly related to the slowness of the scaling in space and time.
Our results and methods, which we expect to be applicable to more general types of models, could open a long-sought path to determining the universality classes behind domain coarsening and phase-ordering kinetics from first principles, which are usually modelled by phenomenological models in near-equilibrium settings.

\begin{acknowledgments}
The authors thank S.~Bartha, J.~Berges, A.~Chatrchyan, Y.~Deller, J.~Dreher, K.~Geier, P.~Gro{\ss}e-Bley, M.~Karl, S.~Lannig, I-K. Liu, M.K.~Oberthaler, J.M. Pawlowski, A.~Pi{\~n}eiro Orioli, M.~Pr\"ufer, N.~Rasch, I.~Siovitz, H.~Strobel, and S.K.~Turitsyn for discussions and collaboration on related topics. 
The authors acknowledge support 
by the ERC Advanced Grant EntangleGen (Project-ID 694561), 
by the German Research Foundation (DFG), through 
SFB 1225 ISOQUANT (Project-ID 273811115), 
grant GA677/10-1, 
and under Germany's Excellence Strategy -- EXC 2181/1 -- 390900948 (the Heidelberg STRUCTURES Excellence Cluster), 
by the state of Baden-W{\"u}rttemberg through bwHPC and DFG through
grant INST 35/1134-1 FUGG (MLS-WISO cluster), 
and grant INST 40/575-1 FUGG (JUSTUS 2 cluster).
A.~N.~M. acknowledges financial support by the IMPRS-QD (International Max Planck Research School for
Quantum Dynamics).\\
\end{acknowledgments}

\clearpage
\begin{appendix}
\widetext
\begin{center}
\textbf{APPENDIX}
\end{center}
\renewcommand{\theequation}{A\arabic{equation}}
\setcounter{equation}{0}
\setcounter{table}{0}
\makeatletter
%

In the following we provide further details of the numerical methodology and results. 
We furthermore give some more information concerning the analytical approach, in particular on performing the scaling analysis which leads to the exponents \eqref{eq:exponents} character\bae{is}{iz}ing the non-thermal fixed point behind coarsening in the sine-Gordon model, cf.~also \cite{Heinen2022b}.

\section{Simulations of sine-Gordon coarsening in the non-relativistic regime}
\label{sec:NumSG}
In this first section, we provide more details underlying our numerical results on universal scaling dynamics of the sine-Gordon model, which we compare with the analytical predictions of \cite{Heinen2022b}, cf.~Eqs.~\eqref{eq:scalrel}, \eqref{eq:porodlaw}, \eqref{eq:exponents}, and \Sect{scalingSG}. 
In our simulations, we consider the low-energy, non-relativistic regime of the sine-Gordon excitations, in analogy to the analytical approximations. 
We first discuss the derivation of the respective equation of motion, which has the form of a non-linear Schr\"odinger equation with Bessel function non-linearity, which we will call Gross-Pitaevskii-Bessel equation  (GPBE) as it is of Gross-Pitaevskii type with the interaction term involving a Bessel function instead of only a cubic non-linearity.
At lowest algebraic order in the argument of the Bessel function, the equation is equivalent to the GPE. 
We then compute, starting from initial states far from equilibrium and using the Truncated Wigner method of semi-classical statistical simulations, the relaxation dynamics according to the GPB equation in two and three dimensions.
As reported in the main text, the systems are found to approach a non-thermal fixed point, exhibiting universal scaling dynamics in space and time, corresponding to coarsening of the field pattern in position space.

\subsection{Sine-Gordon Model}
\label{app:SGmodel}
The `sine-Gordon' equation, which is a non-linear Klein-Gordon equation with a sine-function non-linearity,
\begin{equation}
	\label{app-eq:sGE}
	\Box\varphi + m^{2} \sin \varphi = 0\,, 	
\end{equation}
can be derived as an Euler-Lagrange equation from the Lagrangian density
\begin{equation}
	\label{eq:SG_lagrangian}
	\mathcal{L}_{\mathrm{sG}} 
	=
	\frac{1}{2\eta}\partial_{\mu} \varphi\,\partial^{\,\mu} \varphi + \lambda \left(\cos\varphi - 1\right), 
\end{equation}
which is typically written in terms of the real coupling parameters $\lambda$ and $\eta$, with $\eta\lambda=m^{2}$.
The parameter $\eta$ has the mass dimension $[\eta]=1-d$ in $d$ spatial dimensions, which is chosen such that the real scalar field $\varphi$ is dimensionless. 
The coupling $\lambda$, with $[\lambda]=1+d$, then sets the strength of the interactions.
Writing the cosine potential as its Taylor series, it follows that the coupling constants of all vertices, from second to arbitrarily high order, are fixed by a single parameter, $m^{2}$.  
Note that one may also rescale the dimensionless field $\varphi$ as $\varphi=\phi\sqrt{\eta}$, such that $\phi$ carries the same dimension as in the Klein-Gordon model. 
The sine-Gordon Lagrangian then reads $\mathcal{L}=(\partial\phi)^{2}/2 + \eta^{-1}m^{2}[\mathrm{cos}(\sqrt\eta\phi)-1]$, which shows that $\eta$ controls the relative weight of the higher-order vertices, e.g., of the standard $\lambda_{4}\phi^{4}/4!$ vertex with coupling constant $\lambda_{4}=-\eta m^{2}=-\eta^{2}\lambda$, as compared with $m^{2}\phi^{2}/2$.

\subsection{Non-relativistic limit of the sine-Gordon equation}
\label{sec:NRLimitSG}
In order to derive the non-relativistic limit of the sine-Gordon equation, we follow the standard approach for the derivation of the Gross-Pitaevskii equation (GPE) as the non-relativistic limit of the $\lambda\phi^4$ Klein-Gordon model, cf., e.g.,  \cite{Benson1991a.PhysRevD.44.2480,Evans:1995yz,Berges:2014xea}, which needs to be adapted in the  interaction term only.
We write the sine-Gordon equation derived from \eq{SG_lagrangian} in the form
\begin{align}
\label{eq:EOMrelSineGordon}
	\Box\varphi+m^2\varphi+\eta\lambda\left(\sin\varphi-\varphi\right)=0
	\,,
\end{align}
with $m^2=\eta\lambda$ as defined before. 
The non-relativistic limit is reached for plane-wave momenta $p\ll m$ and associated excitation energies $\omega(\mathbf{p})\approx m+p^2/(2m)$.
The non-relativistic model results in the limit where particle number is well conserved, i.e., imposing a chemical potential on the order of the rest mass $m$.
Hence, the energy of excitations above the non-relativistic ground states is conveniently measured with respect to the energy zero at $\omega(0)=m$. 
Shifting the energy zero implies factoring out fast oscillations with frequency $m$, for which we define the non-relativistic complex field $\psi\in\mathbb{C}$ through
\begin{align}
	\label{eq:representationphi}
	\varphi=\Re\{\psi\exp(-\mathrm{i}mt)\}
	\,.
\end{align}
As a consequence, $\psi$ is a comparatively slowly oscillating field, which obeys
$\left|{\partial^2\psi}/{\partial t^2}\right|\ll m\left|{\partial\psi}/{\partial t}\right|$.
Inserting  \Eq{representationphi} into the sine-Gordon equation \eq{EOMrelSineGordon} and neglecting second order derivatives with respect to time yields
\begin{align}
	\label{eq:eomsinegordon1} 
	&\Re\left[\left(-2\mathrm{i}m\partial_{t}\psi-\Delta \psi \right)\exp(-\mathrm{i}mt)\right] 
	+\eta\lambda\left(\sin\left\{\Re\left[\psi\exp(-\mathrm{i}mt)\right]\right\}
	  -\Re\left[\psi\exp(-\mathrm{i}mt)\right]\right)=0
	\,.
\end{align}
The potential occuring in the interaction term can be rewritten as
\begin{align}
	\label{eq:potterm}
	\sin\{\Re[\psi\exp(-\mathrm{i}mt)]\}-\Re[\psi\exp(-\mathrm{i}mt)] 
	&=\sum_{n=1}^\infty\frac{(-1)^n}{(2n+1)!}\left\{\Re[\psi\exp(-\mathrm{i}mt)]\right\}^{2n+1}
	\nonumber\\
	&=\sum_{n=1}^\infty\frac{(-1)^n}{(2n+1)!}\frac{1}{2^{2n+1}}\sum_{k=0}^{2n+1}\binom{2n+1}{k}\psi^{2n+1-k}(\psi^*)^k 
	\exp\left[-\mathrm{i} \left(2n+1-2k \right)mt \right]
	\,.
\end{align}
To neglect particle-number changing processes, we drop all terms oscillating with a frequency larger than $m$, i.e., we keep only those with $k=n$ and $k=n+1$.
The last line in \Eq{potterm} then reduces to
\begin{align}
	\sum_{n=1}^\infty\frac{(-1)^n}{(2n+1)!}\frac{1}{2^{2n+1}}\frac{(2n+1)!}{n!(n+1)!}|\psi|^{2n}2\Re[\psi\exp(-\mathrm{i}mt)]
	&= \left[\frac{2}{|\psi|}\sum_{n=0}^\infty\frac{(-1)^n}{n!(n+1)!}\left(\frac{|\psi|}{2}\right)^{2n+1}-1\right]
	\Re[\psi\exp(-\mathrm{i}mt)]
	\nonumber\\
	&= \left[\frac{2}{|\psi|}J_1(|\psi|)-1\right]\Re[\psi\exp(-\mathrm{i}mt)]
	\,.
\end{align}
Here, $J_1(x)$ is the Bessel function of the first kind. 
Inserting the above results into \Eq{eomsinegordon1} we find 
\begin{align}
	\Re\left(\left\{-2\mathrm{i}m\frac{\partial\psi}{\partial t}-\Delta \psi 
	+\eta\lambda\left[\frac{2}{|\psi|}J_1(|\psi|)-1\right]\psi\right\}
	\exp[-\mathrm{i}mt]\right)=0
	\,.
\end{align}
This results in the non-linear Schr\"odinger equation of motion of the sine-Gordon model in the non-relativistic limit \cite{Eby:2014fya,Braaten:2016kzc,Robson2021a.arXiv210504498R}
\begin{align}
	\label{app-eq:GPBE}
	\mathrm{i}\partial_{t}\psi
	=-\frac{1}{2m}\Delta \psi +\frac{\eta\lambda}{m}\left(\frac{1}{|\psi|}J_1(|\psi|)-\frac{1}{2}\right)\psi
	\,.
\end{align}
We will refer to \eqref{app-eq:GPBE} as Gross-Pitaevskii-Bessel (GPB) equation.
The associated Lagrangian is given by
\begin{align}
	\label{eq:GPBLagr}
	\mathcal{L}_{\text{GPB}}
	=&\ \frac{\mathrm{i}}{2\eta^r}(\psi^*\partial_t\psi-\psi\partial_t\psi^*)
	-\frac{1}{2m\eta^r}\nabla\psi^*\cdot\nabla\psi
	+ \frac{2\lambda}{m\eta^{r-1}}\left[J_0(|\psi|)+\frac{1}{4}|\psi|^2-1\right]
	\,,
\end{align}
where again the overall factor $\eta^{-r}$, with exponent $r=d/(d-1)$ for $d\not=1$, is chosen such that the field $\psi$ is dimensionless.
Only for $d=1$, one has $r=0$ since $\eta$ as defined for the sine-Gordon model is dimensionless, while the GPB Lagrangian rather requires the multiplication with a constant of mass dimension $1$, e.g., $m$.
Note, that, in any dimension, there is still only a single independent coupling parameter $m$ quantifying the model, as $\eta$ can be absorbed by means of a rescaling of space and time.
In the limit $|\psi|\ll1$, the Lagrangian \eq{GPBLagr} reduces to that of the GP model plus $\mathcal{O}(|\psi|^{6})$ interactions, with GP coupling $g=m/16$. 

\begin{figure}
\centering
	\includegraphics[width=0.49\columnwidth]{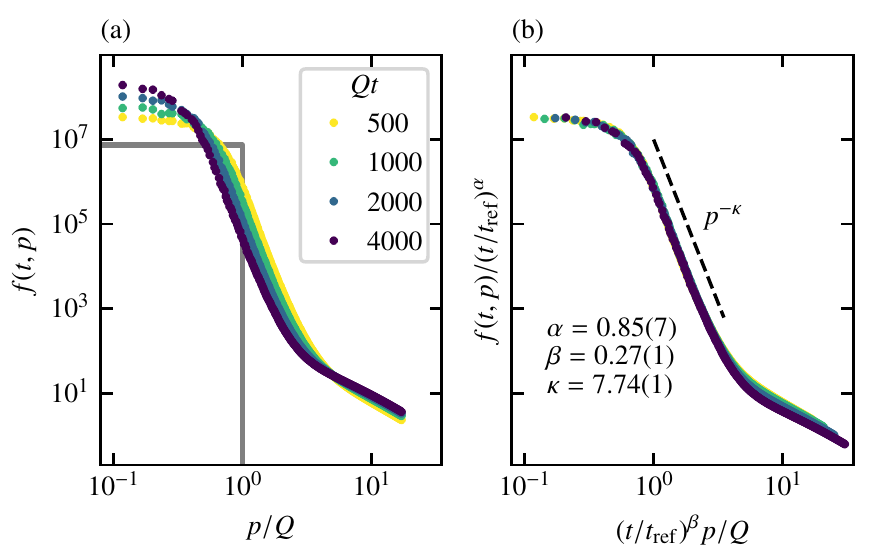}
	\includegraphics[width=0.49\columnwidth]{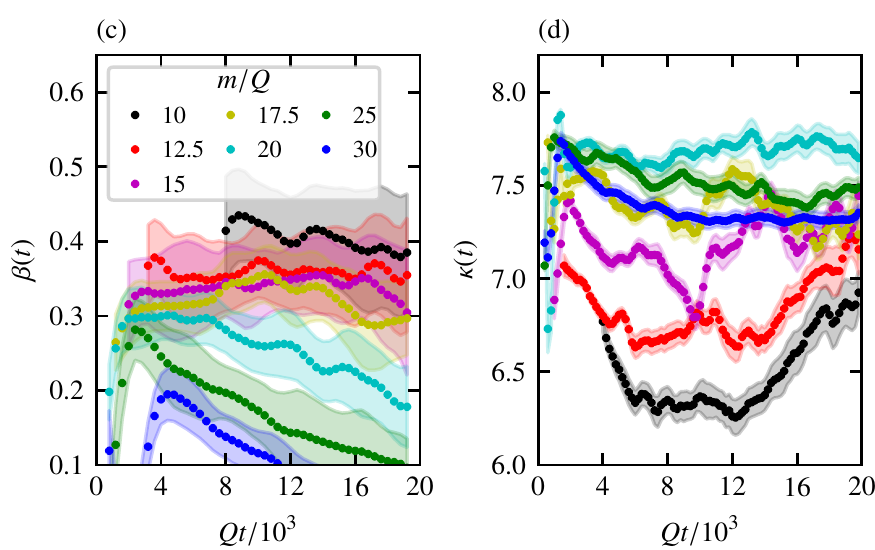}
	\caption{(a) Evolution of the occupation number in the three-dimensional GPB system, for $m/Q=20$, $F_{0}=10^{3}$, averaged over $15$ runs. 
	(b) shows the occupation number for the same four times after rescaling them with the scaling exponents $\alpha=0.85(7)$ and $\beta=0.27(1)$,  with respect to $Qt_{\mathrm{ref}}/10^{3}=0.5$. 
	In the power-law region the spectrum falls off as $\sim p^{-\kappa}$, with $\kappa=7.74(1)$.
	(c) Scaling exponent $\beta$, for $d=3$, as a function of time $t$, obtained by performing a scaling collapse of $f(t,p)$ (each averaged over $12$ runs) at times $t/2$, $3t/4$, $t$. 
	The resulting evolution is shown for $7$ different values of the gap parameter $m/Q$, as given in the legend, and has been smoothed by means of a Savitzky-Golay filter, while the fluctuations are indicated by shaded color bands.
	(d) The same for $\kappa(t)$, obtained by fitting a function of the form \eqref{eq:fMK} to $f(t,p)$, within a range of momenta from the lowest available $p$ to a momentum half-way between the two bending scales.
	Both panels give an indication of scaling with universal exponents for intermediate values of $m/Q$.
	}
	\label{fig:OccupationNumber3D}
\end{figure}

\begin{figure*}
	\includegraphics[width=0.49\columnwidth]{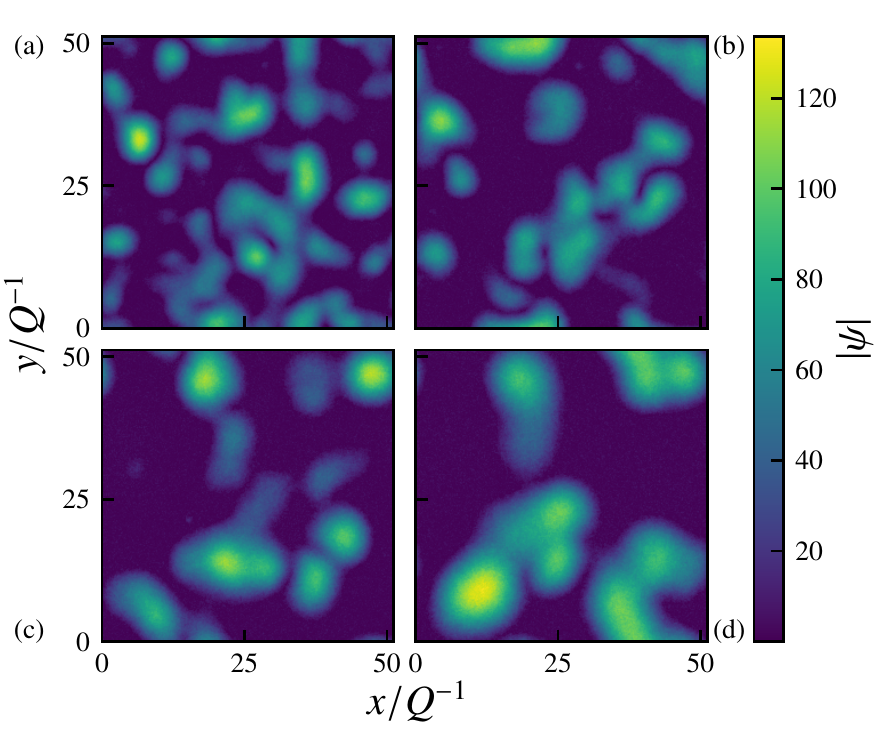}
	\includegraphics[width=0.49\columnwidth]{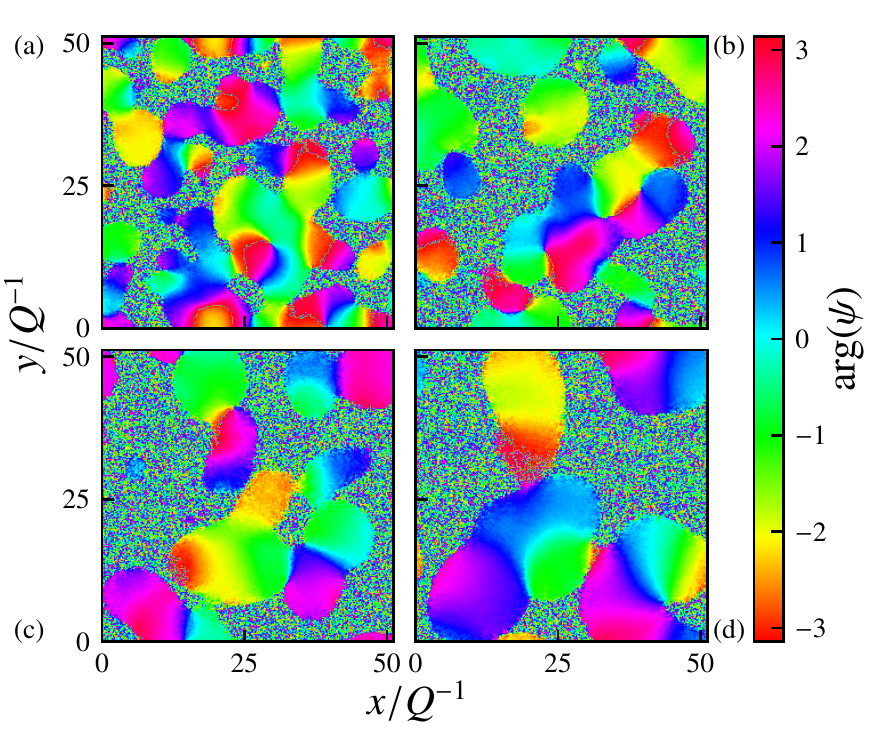}
	\caption{Amplitude $|\psi(\mathbf{x},t)|$ (left panel) and phase angle $\arg[\psi(\mathbf{x},t)]$ (right panel) of the time evolving field distribution in position space for a single run of the simulations in $d=3$ dimensions. 
	The panels show planar slices $\mathbf{x}=(x,y)$ through the 3D system at $z=0$.
	The times of the snapshots in panels (a-d) are the same as those chosen in \Fig{OccupationNumber3D}. 
	Whereas the amplitude remains in the zeroth minimum of the Bessel function at most positions, it grows large within small separated patches. 
	While the phase angle is approximately constant within one patch, it varies randomly between the patches and is deformed during mergers.
	These patches merge and grow over time, giving rise to universal coarsening \bae{behaviour}{behavior} of the system which manifests itself as rescaling in space and time.
	(See \cite{Videos} for videos of exemplary runs.)
	}
	\label{fig:RealSpace3D}
\end{figure*}

\subsection{Universal scaling dynamics}
\label{sec:NumResults} 
To study the universal scaling dynamics of the non-relativistic limit of the sine-Gordon model, we numerically solve the GPB equation \eqref{app-eq:GPBE} by means of a split step Fourier method.
We study the time evolution on a two-dimensional spatial lattice with $2048^2$ grid points and on a three-dimensional lattice with $256^3$ grid points, in both cases subject to periodic boundary conditions.
We set the lattice spacing to unity and express all quantities in the respective numerical units. 

To incorporate fluctuations beyond mean-field, we make use of the Truncated Wigner method \cite{Blakie2008a,Polkovnikov2010a}, choosing the initial field configurations with random phase noise.
Specifically, as for the initial conditions leading to universal scaling dynamics of the Gross-Pitaevskii model \cite{Schole:2012kt,Nowak:2012gd,Orioli:2015dxa}, we take the initial distribution of the occupation number $\langle \lvert \psi(\mathbf{p},0) \rvert^2\rangle\equiv f(t=0,\lvert \mathbf{p}\rvert)$ to be constant up to some momentum cutoff, with the amplitude being subject to Gaussian noise corresponding to half a particle per mode.
The phases $\theta(\mathbf{p},0)$ of the complex field are chosen randomly on the circle. 
Such an initial state is represented by the following density matrix 
\begin{align}
	\label{initdens}\hat{\rho}_0
	=\prod_{|\mathbf{p}|\leq Q}\hat{\rho}^{(\mathbf{p})}
	\prod_{|\mathbf{p}|>Q}|\,0_{\mathbf{p}}\rangle\,\langle0_{\mathbf{p}}\,|
	\,,
\end{align}
with $Q$ being the momentum cutoff of the box distribution, $|\,0_{\mathbf{p}}\rangle$ the vacuum state of momentum mode $\mathbf{p}$, and 
\begin{align}
	\hat{\rho}^{(\mathbf{p})}
	=\frac{1}{2\pi}\int \limits_{-\pi}^\pi \mathrm{d}\theta 
	\left|\sqrt{f(0,\lvert \mathbf{p}\rvert)}e^{i\theta}\right\rangle
	\left\langle \sqrt{f(0,\lvert \mathbf{p}\rvert)}e^{i\theta}\right|
	\,.
\end{align}
Here, $|\alpha\rangle$ denotes a coherent state with complex eigenvalue $\alpha$, $|\alpha|^{2}=f(0,\mathbf{p})$, of the momentum-space field operator $\hat\psi(\mathbf{p})$. 

As parameters we choose $\eta=1$, $\rho_0 \equiv N_0/\mathcal{V} = 10^3$ for the simulations in both, $d=2$ and $d=3$ dimensions, on a lattice with volume $\mathcal{V}$. 
We set $Q=0.05$, $m=\sqrt{\eta\lambda}=1$ in two dimensions, and $Q=0.2$, $m=\sqrt{\eta\lambda}=4$ in three dimensions.
The particle number is chosen such that the initial field amplitude $\lvert \psi_0\rvert\gg1$ is sufficiently large to be far away from the Gross-Pitaevskii limit of the GPB equation. 
When equally spread across all lattices sites, the initial average field amplitude corresponds to a mean density $\lvert \psi_0\rvert=\sqrt{\rho_0} \approx31.6$, which is close to the sixth minimum of the Bessel function.

\begin{figure}
	\centering
	\includegraphics[width=0.49\columnwidth]{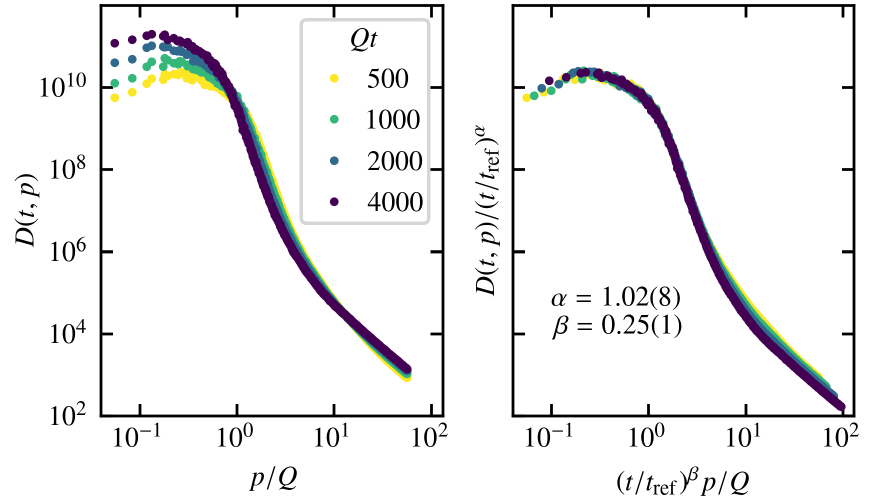}
	\includegraphics[width=0.49\columnwidth]{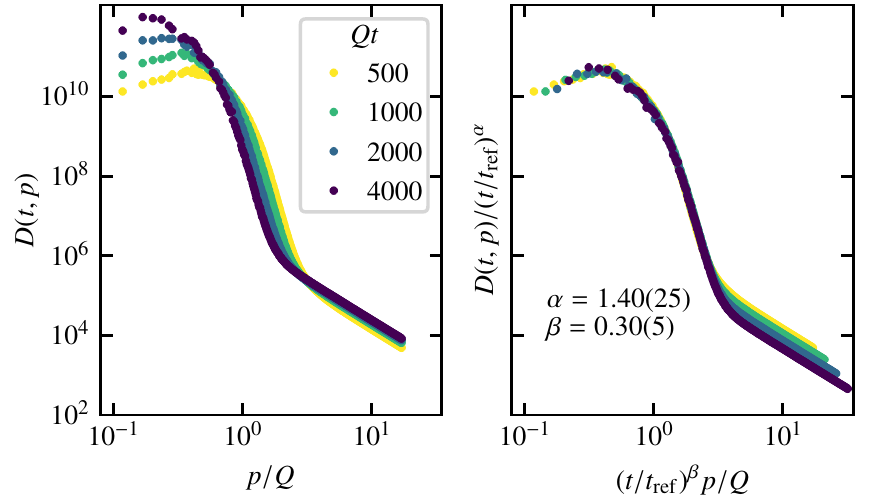}
	\caption{Evolution of the density-density correlator in the two-dimensional GPB system (left two panels) as well as in three dimensions (right two panels). In 2D, we observe self-similar scaling in time with $\alpha=1.02(8)$ and $\beta=0.25(1)$, in 3D with $\alpha=1.40(25)$ and $\beta=0.30(5)$. Since $\beta$ takes a similar value as for the occupation number spectrum, one may suspect that it is the merging of high density blobs that drives the scaling.
	}
	\label{fig:DensityDensity2D3D}
\end{figure}
The time evolution of the GPB equation \eqref{app-eq:GPBE} is then performed with a time stepping, in numerical units, of $\delta t=0.02$, corresponding to $\delta t =0.001\,Q^{-1}$ in $d=2$ dimensions and $\delta t =0.004\,Q^{-1}$ in $d=3$.

The evolving occupation number distribution $f(t,\mathbf{p})$ is averaged, on the cubic momentum grid, over the angular orientations $\mathbf{p}/p$ as well as over 25 ($d=2$) and 15 ($d=3$) real\bae{is}{iz}ations. 
The resulting radial distributions $f(t,p)$ are shown, for the $d=2$ case with $m/Q=20$, in Fig.~\ref{fig:OccupationNumber2D} (left panel) in the main text, and, for $d=3$ with $m/Q=20$, in \Fig{OccupationNumber3D}(a) above.

For times   $Qt\gtrsim500$, we observe a self-similar evolution of the occupation number according to the scaling relation 
\begin{equation}
	\label{appeq:scalrel}
	f(t,p)  = \left (t/t_{\mathrm{ref}} \right)^{\alpha} f_s \left ( \left [ t/t_{\mathrm{ref}} \right]^{\beta} p \right ),
\end{equation} 
with scaling function $f_s(p)$, scaling exponents $\alpha$, $\beta$, and reference time, which we set to  $t_{\mathrm{ref}}=500/Q$ at the beginning of the temporal scaling regime, cf.~Figs.~\ref{fig:OccupationNumber2D} (right panel) and \Fig{OccupationNumber3D}(b).
By means of a least-square algorithm we extract the scaling exponents
\begin{align}
	\alpha&=0.47\pm 0.08\,,\quad
	\beta=0.23\pm 0.01\,; &(d=2)
	\\	
	\alpha&=0.85\pm 0.07\,,\quad
	\beta=0.27\pm 0.01\,. &(d=3)
\end{align}
The exponents $\kappa$ are found by fitting the power law \eqref{eq:porodlaw} to the numerical data for $f(t,p)$, within a range of momenta from the lowest available $p$ to a momentum half-way between the two bending scales.
The numerically found values are
\begin{align}
	\kappa&=5.68\pm 0.01\,; &(d=2)
	\\	
	\kappa&=7.74\pm 0.01\,. &(d=3)
\end{align}
For the case $d=2$, our numerical data rather well corroborates our analytical prediction $\beta=1/(2+d)$, $\alpha=d\beta$,  $\kappa=2d+2$, cf.~Eqs.~\eqref{eq:exponents} and \eq{alphabetaanomalous}, while for $d=3$, the deviations are stronger.
Nevertheless, both exponents indicate  scaling with exponents $\beta$ ($\kappa$), which are substantially smaller (larger) than the standard $\beta=1/2$ ($\kappa=d+1$), see \Eq{alphabetagaussian}. 
The observed sensitivity could be due to the condition that the predicted exponents rely on exponentially large occupancies, see \cite{Heinen2022b}.

Despite the small statistical errors of the above exponents, we found a rather strong sensitivity of $\alpha$ and $\beta$ on the initial condition, in particular on the ratio of the momentum cutoff of the box $Q$ and the mass $m$.
Moreover, the extracted scaling exponents showed temporal variations, depending on these parameters.
As described in the main text, we estimated these effects by evaluating the self-similarity between time steps $t/2$, $3t/4$, and $t$ as a function of $t$, for different choices of $m/Q$, keeping $Q$ constant.
The resulting time dependence of the exponents $\beta$ and $\kappa$ are shown, for $d=2$, in Fig.~\ref{fig:betakappaoft2D} in the main text and, for $d=3$, in Figs.~\fig{OccupationNumber3D}(c,d) above.
The data has been smoothed, in time, by means of a Savitzky-Golay filter, which fits the data with a polynomial of order $3$ on ($d=2$) $31$ and ($d=3$) $7$ equally spaced points along the time axis,
and the temporal fluctuations are indicated by the shaded colored bands.

To provide more insight into the physical mechanism underlying the universal scaling evolution, we depict snapshots of the position-space configurations of a single run for both, the two- and three-dimensional systems:  
As in Fig.~\ref{fig:RealSpace2D} in the main text, in \Fig{RealSpace3D}, the left set of panels shows the field amplitude $|\psi(\mathbf{x},t)|$, for two-dimensional planar slices through the 3D volume, at the four times during the scaling period chosen at which the spectra were shown in \Fig{OccupationNumber3D}, while the right set depicts the respective phase angles $\theta(\mathbf{x},t)=\mathrm{arg}[\psi(\mathbf{x},t)]$ of the same field distributions.
See also \cite{Videos} for videos of exemplary runs, where the energy of the field has been renormal\bae{ise}{ize}d by subtracting the constant $-\eta\lambda\psi/(2m)=-m\psi/2$ from the e.o.m.~\eqref{app-eq:GPBE}, such that the phase of the field is constant in time where $\psi$ is spatially uniform.

In the left panels of these figures, showing $|\psi(\mathbf{x},t)|$, we observe clearly separated spatial patches where $\lvert \psi \rvert$ is large as compared with its value elsewhere, where it resides in the 0th minimum of the Bessel function.
The observed pattern can be attributed to the fact that, on the one hand, the potential $-J_0(\lvert\psi\rvert)$ has an absolute minimum at $\lvert\psi \rvert=0$ while, on the other, it approaches $0$ for $\lvert\psi \rvert \to\infty$.
Recall that the Bessel function for large $x$ asymptotically assumes the form $J_0(|\psi|)\sim\sqrt{2/\pi |\psi|}\cos(|\psi|-\pi/4)$.

Patches of large field amplitude merge over time and thus form larger patches.
This gives rise to coarsening evolution of the field pattern, which manifests itself in spatio-temporal scaling evolution of the occupation number distribution $f(t,p)$.
Since the characteristic scale $p_{\Lambda}$, at which the scaling form $f_\mathrm{s}(p)$ changes from its plateau value at smaller $p$ to the $\sim p^{-\kappa}$ fall-off, is a measure for the inverse mean size of the patches, its temporal rescaling $p_{\Lambda}(t)\sim t^{-\beta}$ describes power-law coarsening of the characteristic length scale.

This picture is further corroborated by considering the equal-time density-density correlator $D(t,p)$, which is defined as the Fourier transform of $\langle |\psi(\mathbf{x},t)|^2|\psi(\mathbf{y},t)|^2\rangle$ with respect to $\mathbf{x}-\mathbf{y}$ and angle averaged over the orientation of $\mathbf{p}$.
$D(t,p)$ is shown, for the same times as before, in \Fig{DensityDensity2D3D}, for $d=2$ and $3$, respectively. 
Since the resulting coarsening exponents $\beta$ are close to those of the occupation number spectra $f$, the data suggests that the density fluctuations are dominating the spatio-temporal scaling. 

\subsection{Non-locality of the transport}
\label{sec:nonlocality}
Our results suggest that a distinct difference between coarsening and standard wave turbulence is their degree of locality in momentum space.
(Wave) turbulence is generically dominated by (near-)local scattering in momentum space which renders the dynamics in the inertial interval \eqref{eq:porodlaw} independent of the physics at its ultraviolet (UV) and infrared (IR) ends \cite{Zakharov1992a,Nazarenko2011a,Balk1988a,Balk1990a,Balk1990b}.
Within an isotropic cascade, this is well described by a continuity equation, 
\begin{equation}
	\label{eq:MomentumSpaceContinuityEq}
	0=\partial_{t}N_\mathrm{Q}(t,p)+\partial_{{p}}{J}_\mathrm{Q}(t,p)
\end{equation} 
for the radial quasiparticle number $N_\mathrm{Q}(t,p)=\Omega_{d}p^{d-1}f(t,p)$ and radial current ${J}_\mathrm{Q}$ \cite{Zakharov1992a}.
Fully developed wave turbulence implies that, within the inertial range of the cascade, the distribution is stationary, $\partial_{t}N_\mathrm{Q}(t,p)=0$. 
In this case, the net flow into and out of each momentum shell $p$ vanishees within the inertial range, quantified by a uniform constant radial current ${J}_\mathrm{Q}(t_{0},p)=\,$const.

In contrast to this, the quasiparticle distribution is not constant in our case, within the regime of momenta where the scaling function falls off as $\sim p^{-\kappa}$.
As can be read off the scaling form \eqref{eq:fMK}, which changes in time according to $p_{\Lambda}(t)\sim t^{-\beta}$, the radial current, as a consequence of the transport equation \eq{MomentumSpaceContinuityEq} rescales in time as
\begin{equation}
	\label{eq:RadialCurrentfromMKf}
	{J}_\mathrm{Q}(t,p)
	\sim t^{-1}\beta\,\Omega_{d} \int_{p/p_{\Lambda}}^{\infty}\mathrm{d}x\,\frac{x^{d-1}}{1+x^{\kappa}}
	\left(d-\frac{\kappa}{1+x^{-\kappa}}\right)
	\,.
\end{equation} 
To get an estimate of the asymptotic \bae{behaviour}{behavior} of the radial current \eq{RadialCurrentfromMKf}, we note that  the exponents \eqref{eq:exponents} imply that
\begin{align}
	{N}_\mathrm{Q}(t,p)
	\sim \left\{\begin{array}{ll}
	t^{d/(d+2)}p^{d-1}\sim t^{1/(d+2)} (p/p_{\Lambda})^{d-1}\,,
	&\text{for}\quad p\ll p_{\Lambda}\\
	t^{-1}p^{-d-3}\sim t^{1/(d+2)} (p/p_{\Lambda})^{-d-3}\,,
	&\text{for}\quad p\gg p_{\Lambda}
	\end{array}\right.
	\,,
\end{align} 
with $p\ll p_{\lambda}$ implied.
The radial current \eq{RadialCurrentfromMKf}, on the other hand, scales as 
\begin{align}
	{J}_\mathrm{Q}(t,p)
	\sim (p/t)N_{\mathrm{Q}}(t,p)
	\sim\left\{\begin{array}{ll}
	t^{-2/(d+2)}p^{d}\sim t^{-1} (p/p_{\Lambda})^{d}\,,
	&\text{for}\quad p\ll p_{\Lambda}\\
	t^{-2}p^{-d-2}\sim t^{-1} (p/p_{\Lambda})^{-2-d}\,,
	&\text{for}\quad p\gg p_{\Lambda}
	\end{array}\right.
	\,.
\end{align} 
As a result, the radial current is peaked around the characteristic scale $p\approx p_{\Lambda}$.
Other than for local cascades of stationary wave-turbulent flows, the current decreases in time, and, like the distribution in the power-law regime \eqref{eq:porodlaw}, it depends strongly on the location of the IR cutoff scale $p_{\Lambda}$.
Hence, the process underlying anomalous coarsening according to the sine-Gordon model is in a regime, where the redistribution is strongly nonlocal in momentum space: 
Since the quasiparticle number is conserved in the elastic collisions, a (strongly) non-uniform current means that there is considerable momentum transfer in these collisions, which is a necessary precondition for accumulating occupancy in certain modes more strongly than others get depleted. 
Fig.~\ref{fig:SourcesJQ} depicts the momentum distribution of the temporally rescaled current $t(d\beta\Omega_{d})^{-1} J_{Q}$, \Eq{RadialCurrentfromMKf}. 

In the scaling evolution found here, the modes around, i.e., just below $p_{\Lambda}$ undergo the strongest growth while all modes $p_{\Lambda}\lesssim p\lesssim p_{\lambda}$ get depleted, the more the closer they are to $p_{\Lambda}$.
In fact, all modes $p\lesssim p_{\Lambda}$ grow, but due to the volume factor, most of the quasiparticles are deposited closer to $p_{\Lambda}$ and demonstrates its concentration around the scale $p\approx p_{\Lambda}$.

Note, finally, that this can be compared with the situation at the `Gaussian' non-thermal fixed point, which is defined by the scaling with $\beta=1/2$, $\alpha=d\beta$, and $\kappa=d+1$ \cite{Orioli:2015dxa,Walz:2017ffj.PhysRevD.97.116011,Chantesana:2018qsb.PhysRevA.99.043620} and which entails a self-similar buildup of an out-of-equilibrium quasicondensate from incoherent phase excitations \cite{Mikheev:2018adp}:
This fixed point is also characterised by a nonuniform radial transport decaying in time, viz.~${J}_\mathrm{Q}(t,p)\sim t^{-1/2}p^{-1}\sim t^{-1}(p/p_{\Lambda})^{-1}$ for $p\gg p_{\Lambda}$ within the $p^{-\kappa}$ tail of $f$, independent of $d$.
For $p\ll p_{\Lambda}$ it scales like the anomalous one, as ${J}_\mathrm{Q}(t,p)\sim t^{-(d+2)/2}p^{d}\sim t^{-1}(p/p_{\Lambda})^{d}$.
Hence, while the Gaussian fixed point is also character\bae{ise}{ize}d by a non-local current, the anomalous one is more strongly peaked at the IR cutoff scale and shows a distinctly steeper power-law decay at momenta larger than $p_{\Lambda}$, as illustrated in Fig.~\ref{fig:SourcesJQ}.
The associated non-locality of the transport in momentum space is further corroborated by the analytical results summar\bae{ise}{ize}d  in the following, see \Sect{nonlocality-in-kinetics}.

\section{\label{sec:scalingSG}Universal scaling according to sine-Gordon kinetics}
In this section, we briefly sketch the derivation of the scaling properties close to the anomalous non-thermal fixed point as predicted for the sine-Gordon model in the low-energy limit \cite{Heinen2022b}.

\subsection{Scattering integral and $T$-matrix}
\label{sec:ScattInt}
Using a non-perturbative approach based on functional field theoretic techniques, the scattering integral $C[f]$, Eq.~\eqref{eq:Cfmain2}, results as a sum of multidimensional integrals over the spatial momenta, involving a scattering `$T$-matrix', energy- and momentum-conservation constraints, and a sum of in- and out-scattering terms depending on the quasiparticle distribution $f(t,\mathbf{p})$ only, 
\begin{align}
	\label{app-eq:Cfmain2}
	&C[f](t,\mathbf{p})
	\equiv
	\sum_{n=1}^{\infty}C^{(n)}[f](t,\mathbf{p})
	=-\sum_{n=1}^{\infty}
	\int\prod_{i=1}^{2n+1}\frac{\dd{\mathbf{q}_{i}}}{(2\pi)^{d}} \,
	\left|T^{(n)}(t;{\mathbf{p},\mathbf{q}_{1},\dots,\mathbf{q}_{2n+1}})\right|^{2}
	\nonumber\\
	&\quad\times\
	\delta(\omega_{\mathbf{p}}-\omega_{\mathbf{q}_{1}}-\dots-\omega_{\mathbf{q}_{n+1}}
	+\omega_{\mathbf{q}_{n+2}}+\dots+\omega_{\mathbf{q}_{2n+1}})\,
	\delta(\mathbf{p}-\mathbf{q}_{1}-\dots-\mathbf{q}_{n+1}+\mathbf{q}_{n+2}+\dots+\mathbf{q}_{2n+1})
	\nonumber\\
	&\quad\times\
	\left[(f_{\mathbf{q}_{1}}+1)\cdots(f_{\mathbf{q}_{n+1}}+1)f_{\mathbf{q}_{n+2}}\cdots f_{\mathbf{q}_{2n+1}}f_{\mathbf{p}}
	\right.
	-\left.
	f_{\mathbf{q}_{1}}\cdots f_{\mathbf{q}_{n+1}}(f_{\mathbf{q}_{n+2}}+1)\cdots 
	(f_{\mathbf{q}_{2n+1}}+1)(f_{\mathbf{p}}+1)\right]
	\,,
\end{align}
where, on the right-hand-side, the dependence of $f_{\mathbf{q}}\equiv f(t,\mathbf{q})$ on the time $t$ is suppressed.
Universal transport is dominated by the infrared wave numbers below the gap energy, $\omega(|{\mathbf{p}}|\to0)\to M$, such that we need to take into account on-energy-shell terms only.
Hence, the above scattering integral describes $(n+1)$-to-$(n+1)$ processes for which the sum of all frequencies, $p^{0}+\sum_{i=1}^{2n+1}q^{0}_{i}$ is gapless, i.e., $n+1$ of the frequencies are evaluated in the positive domain, $q^{0}_{i}=\omega(\mathbf{q}_{i})$, $i=1,\dots,n+1$, and a further $n+1$ in the negative domain, $q^{0}_{i}=-\omega(\mathbf{q}_{i})$, $i=n+2,\dots,2n+1$, as well as $p_{0}=-\omega(\mathbf{p})$.

The $T$-matrices squared turn out to read
\begin{align}
	\label{app-eq:Tmatrix}
	\left|T^{(n)}(t;{\mathbf{p},\mathbf{q}_{1},\dots,\mathbf{q}_{2n+1}})\right|^{2}
	&=\frac{g_{\mathrm{eff}}^{2}(n;t;\mathbf{p},\{\mathbf{q}_{i}\})}{n!(n+1)!}
	\frac{\eta^{2n+2}}{2\omega(\mathbf{p})}\prod_{i=1}^{2n+1}\frac{1}{2\omega(\mathbf{q}_{i})}
	\,,
\end{align}
in terms of the effective coupling function $g_{\mathrm{eff}}$,
\begin{align}
	\label{eq:SM:geff}
	g_{\mathrm{eff}}^{2}(n;t;\mathbf{p},\{\mathbf{q}_{i}\})
	&=\sum_{m=1}^{n}\left[
	\sum_{\{\sigma\}}\left|\Lambda^{R}_{\mathrm{e}}
	\left(t,
	\sum_{i=1}^{2m}s(\sigma_{i})\,\omega_{\mathbf{q}_{\sigma_{i}}},
	\sum_{i=1}^{2m}s(\sigma_{i})\,\mathbf{q}_{\sigma_{i}}
	\right)\right|^{2}\,\right.
	+\left.
	\sum_{\{\sigma\}}\left|\Lambda^{R}_{\mathrm{o}}
	\left(t,
	\sum_{i=1}^{2m+1}s(\sigma_{i})\,\omega_{\mathbf{q}_{\sigma_{i}}},
	\sum_{i=1}^{2m+1}s(\sigma_{i})\,\mathbf{q}_{\sigma_{i}}
	\right)\right|^{2}\,\right]
	\,.
\end{align}
Here, $s(k)=\mathrm{sgn}(n+3/2-k)$, which is $s(k)=+1$ if $k\leq n+1$ and $s(k)=-1$ for $k>n+1$, and the sums over $\sigma\subset\{1,\dots,2n+1\}$ are those over all subsets of $2m$ (or $2m+1$, in the odd case) momenta of all the $\mathbf{q}_{k}$ in a given term.
The nonperturbative coupling functions entering the sum result from a geometric-series-type resummation of loop-chain diagrams in the field-theoretic description involving an even (odd) number of propagators.
They are defined as
\begin{align}
	\label{eq:LambdaReopMain}
	\Lambda^{R}_{\mathrm{e/o}}(t,p)
	&=\frac{\bar\lambda}{1 \mp \bar\lambda\,\Pi_{\mathrm{e/o}}^R(t,p)}
	\,,&
	\bar\lambda 
	&=\lambda\exp\left\{-\frac{\eta}{2}\int_{\mathbf{p}}\frac{f(t,\mathbf{p})+1/2}{\omega_{\mathbf{p}}}\right\}
	\,,
\end{align}	
with a dressed coupling $\bar\lambda$ and
(retarded) loop functions $\Pi^{R}_{e/o}$, which themselves form sums over arbitrarily high powers of distribution functions,
\begin{align}
 	\label{eq:PiReQuasiP}
	&\Pi^R_{\mathrm{e}}(t,p^{0},\vec p) 
  	=-\frac{1}{2\pi}\sum_{n=1}^{\infty}
  	\int 
	\prod_{i=1}^{2n}\frac{\dd{\mathbf{q}_{i}}}{(2\pi)^{d}2\omega(\mathbf{q}_{i})} \,
	\sum_{m=0}^{2n}
	\frac{\eta^{2n}}{m!(2n-m)!}
	\left(p^{0}-\omega_{\mathbf{q}_{1}}-\dots-\omega_{\mathbf{q}_{m}}
	+\omega_{\mathbf{q}_{m+1}}+\dots+\omega_{\mathbf{q}_{2n}}+\mathrm{i}\epsilon\right)^{-1}
	\nonumber\\
	&\times\
	\delta(\mathbf{p}-\mathbf{q}_{1}-\dots-\mathbf{q}_{m}+\mathbf{q}_{m+1}+\dots+\mathbf{q}_{2n})
	\left[(f_{\mathbf{q}_{1}}+1)\cdots(f_{\mathbf{q}_{m}}+1)f_{\mathbf{q}_{m+1}}\cdots f_{\mathbf{q}_{2n}}
	\right.
	-\left.
	f_{\mathbf{q}_{1}}\cdots f_{\mathbf{q}_{m}}(f_{\mathbf{q}_{m+1}}+1)\cdots 
	(f_{\mathbf{q}_{2n}}+1)\right]
	\,,\\ 
 	\label{eq:PiRoQuasiP}
  	&\Pi^R_{\mathrm{o}}(t,p^{0},\vec p) 
  	=-\frac{1}{2\pi}\sum_{n=1}^{\infty}
  	\int 
	\prod_{i=1}^{2n+1}\frac{\dd{\mathbf{q}_{i}}}{(2\pi)^{d}2\omega_{\mathbf{q}_{i}}} \,
	\sum_{m=0}^{2n+1}
	\frac{\eta^{2n+1}}{m!(2n-m+1)!}
	\left(p^{0}-\omega_{\mathbf{q}_{1}}-\dots-\omega_{\mathbf{q}_{m}}
	+\omega_{\mathbf{q}_{m+1}}+\dots+\omega_{\mathbf{q}_{2n+1}}+\mathrm{i}\epsilon\right)^{-1}
	\nonumber\\
	&\times\
	\delta(\mathbf{p}-\mathbf{q}_{1}-\dots-\mathbf{q}_{m}+\mathbf{q}_{m+1}+\dots+\mathbf{q}_{2n+1})
	\left[(f_{\mathbf{q}_{1}}+1)\cdots(f_{\mathbf{q}_{m}}+1)f_{\mathbf{q}_{m+1}}\cdots f_{\mathbf{q}_{2n+1}}
	\right.
	-\left.
	f_{\mathbf{q}_{1}}\cdots f_{\mathbf{q}_{m}}(f_{\mathbf{q}_{m+1}}+1)\cdots 
	(f_{\mathbf{q}_{2n+1}}+1)\right]
  	  \,.
\end{align}
The infinite sums over $n$ in both, the scattering integral and the loop functions, reflect the structure of the cosine potential, which expands to a series of vertices of arbitrarily high power in $\varphi$.

\subsection{Infrared fixed points: Sources of different scaling}
For this, one presupposes, in the late-time scaling limit the scaling form \eqref{appeq:scalrel} of the distribution function.
In the non-relativistic or low-energy limit $p\ll M$, one has
$\omega(\mathbf{p})\approx M+\frac{\mathbf{p}^2}{2M}$,
and energy conservation reduces to that for $\varepsilon(\mathbf{p})={\mathbf{p}^2}/{2M}=s^{-z} \varepsilon\left(s\, \mathbf{p}\right)$, with dynamic exponent $z=2$.
Moreover, all other factors $\omega(\mathbf{q}_i)$ in \eqref{app-eq:Tmatrix}, in the scaling limit, are approximately given by $M$, such that also the dressed coupling $\bar\lambda$ remains constant. 

The scattering integral \eqref{app-eq:Cfmain2} allows identifying a measure for distinguishing regimes which can give rise to different infrared fixed points.
Presupposing that the order of magnitude of the coupling functions $\Lambda^R_{\mathrm{e/o}}$ is roughly equal for all $m$ and all $\sigma$, the sum over the subsets $\sigma$ appearing in \eqref{app-eq:Tmatrix} is approximately proportional to the number of these subsets, such that their sum over $m$ in \eqref{app-eq:Tmatrix} can be estimated to scale as $\approx2^{2n+1}$, for $n\gg1$ \cite{Heinen2022b}.
The momentum integrals over the distribution functions $f_{\mathbf{q}_{i}}$ scale with the quasiparticle density $n_0$, such that the scattering integral scales as
\begin{align}
	C^{(n)}[f]
	\sim \left(\frac{\eta n_{0}}{M}\right)^{2n+1}\frac{1}{n!(n+1)!}
	\sim\frac{F_{0}}{n+1}\left(\frac{F_0^{n}}{n!}\right)^2
	\,,
	\label{eq:SGness}
\end{align}
where $F_0\equiv\eta n_{0}/M$.
Hence, $F_0$ indicates the order $n$ that dominates the collisional integral:  
If $F_0\ll 1$, all terms $C^{(n)}[f]$ beyond $n=1$ can be neglected, and one recovers the standard wave-Boltzmann scattering integral of $\phi^4$ theory \cite{Orioli:2015dxa}. 
In contrast, for $F_0\gg 1$ the sine-Gordon interaction matters, as the order $n\approx F_0\gg1$ dominates the expansion of  the scattering integral, in accordance with the exponential growth of the hyperbolic functions entering the collision integral \eqref{app-eq:Cfmain2}.

\subsection{Spatio-temporal scaling analysis of the kinetic equation}
\label{sec:ScalingAnalysis}
Inserting \eqref{appeq:scalrel} into \eqref{eq:KinEq} and rescaling $\mathbf{p}\to(t/t_{0})^{-\beta}\mathbf{p}$ results in 
\begin{align}
	\label{eq:ScalingKinEq}
	(t/t_{0})^{\alpha-1}\left[\alpha f_\mathrm{s}(\mathbf{p})+\beta\mathbf{p}\cdot\partial_{\mathbf{p}} f_\mathrm{s}(\mathbf{p})\right] 
	=t_{0}\,(t/t_{0})^{-\beta\mu} C[f_\mathrm{s}](\mathbf{p})
	\,.
\end{align}
where $\mu$ defines the scaling dimension of the scattering integral,
$C[f](t,\mathbf{p})=(t/t_0)^{-\beta\mu} C[f_\mathrm{s}]([t/t_0]^\beta \mathbf{p})$.
Hence, if $\alpha-1=-\beta\mu$ is fulfilled, the solution $f$ can assume the scaling form \eqref{appeq:scalrel}.
A second relation between the exponents is obtained from the form \eqref{app-eq:Cfmain2} of the scattering integral.
To find this relation, one considers a single summand $C^{(n)}$ at order $n$ which contains the gain and loss terms
$(f_{\mathbf{q}_{1}}+1)\cdots(f_{\mathbf{q}_{n+1}}+1)f_{\mathbf{q}_{n+2}}\cdots 
f_{\mathbf{q}_{2n+1}}f_{\mathbf{p}}
-f_{\mathbf{q}_{1}}\cdots f_{\mathbf{q}_{n+1}}(f_{\mathbf{q}_{n+2}}+1)\cdots 
(f_{\mathbf{q}_{2n+1}}+1)(f_{\mathbf{p}}+1)$.
For $f_\mathbf{q}\gg1$, which will be the case at low wave numbers $\mathbf{q}$, the leading contribution to these terms contains $2n+1$ factors $f$, while the terms with $2n+2$ such factors cancel each other. 
The dominating terms with $2n+1$ factors contain $2n$ distributions $f_{\mathbf{q}_{i}}$ which are integrated over $\mathbf{q}_{i}$ and one factor $f_{\mathbf{p}}$ which depends on the external momentum.
But there are still $2n+1$ integrals over momenta $\mathbf{q}_{i}$, of which, hence, a single one over $\mathbf{k}\equiv\mathbf{q}_{j}$, $j\in\{1,...,2n+1\}$, is not constrained by a distribution function $f_{\mathbf{q}_{j}}$. 
This single integration, in the limit of large $n$, due to the central-limit theorem, is argued to be unconstrained \cite{Heinen2022b}, such that 
\begin{align}
	\label{eq:CfnofKfn}
	C^{(n)}[f](t,\mathbf{p})
	&\sim-\int \limits^{\Lambda_\text{UV}} d^dk\,
	K^{(n)}[f]\left(t,\mathbf{p};\varepsilon(\mathbf{k}),\mathbf{k}\right)
	f(t,{\mathbf{p}})
	\,,
\end{align}
with the kernel integral being approximately proportional to a Gaussian distribution,
\begin{align}
	\label{eq:KfnGauss}
	&K^{(n)}[f](t,\mathbf{p};E,\mathbf{k})
	\propto  \exp\left(-\frac{E^2}{2\sigma_E^2}\right)\exp\left(-\frac{\mathbf{k}^2}{2\sigma_k^2}\right)
	\,,
\end{align}
where the standard deviations scale as $\sigma_E\sim [M+\varepsilon(k_{\Lambda})]\sqrt{n}$ and $\sigma_k\sim k_{\Lambda}\sqrt{n}$. Here, $k_{\Lambda}$ measures the width $\langle k^{2}\rangle_{f}=\int_\mathbf{k}k^{2}f(t,\mathbf{k})/\int_\mathbf{k}f(t,\mathbf{k})$ of the distribution $f(t,{\mathbf{k}})$ in momentum space, which, 
for the case of the spatial scaling form $f(t,{\mathbf{p}})\sim[p_{\Lambda}^{\kappa}+p^{\kappa}]^{-1}$ found in our numerical simulations, cf.~\Sect{NumSG}, 
is set by the infrared cutoff scale $p_{\Lambda}$, below which the distribution is constant in $p$.
Furthermore, as $p_{\Lambda}\ll M$, which sets a UV cutoff scale $\Lambda_\text{UV}\sim M$,
if $\varepsilon(\Lambda_\text{UV})\ll \sigma_E$ and ${\Lambda}_\text{UV}\ll \sigma_k$, which will be the case for sufficiently large $n\approx F_{0}\gg(\Lambda_{\mathrm{UV}}/p_{\Lambda})^{2}$, then \eq{CfnofKfn} becomes
$C^{(n)}[f](t,\mathbf{p})
	\sim-\Lambda_\text{UV}^d\,
	K^{(n)}[f]\left(t,\mathbf{p};0,0\right)
	f(t,{\mathbf{p}})$.

This dependence of the collisional integral on $f$, together with the full definition of the kernel function $K^{(n)}$ yields the scaling exponent $\mu$ defined above.
Taking, thereby, the integral over $\mathbf{k}$ to be not contributing to the scaling of the scattering integral \eqref{app-eq:Cfmain2} results in
\begin{align}
\label{eq:betamu}
	\mu
	&=2n(d-\alpha/\beta)-d-z+2m-\alpha/\beta
	\,,
\end{align}
where $m$ denotes the scaling exponent of the $T$-matrix,
$|T^{(n)}(t;\mathbf{p},\{\mathbf{q}_{i}\})|
= s^{-m} |T^{(n)}(s^{-1/\beta}t;s\mathbf{p},\{s\mathbf{q}_{i}\})|$.

Finally, since all sums in the series over orders $n$ will have to scale the same way, independence of $n$ requires
$\alpha=d\beta$,
such that 
$\mu=2m-d-z-\alpha/\beta$.
This, in retrospect, also ensures the total quasiparticle density, which is given by the $d$-dimensional momentum integral over $f_{\mathbf{p}}$, to be constant in time.

The scaling of $\Pi^R_\mathrm{e/o}$, Eqs.~\eq{PiReQuasiP}, \eq{PiRoQuasiP}, can be analysed accordingly. 
For high occupancies $f(t,\mathbf{p})\gg1$ in the IR, the loop functions will dominate, 
$\bar{\lambda}\Pi^R_\mathrm{e/o}\gg 1$,
in the denominator of  $\Lambda^R_\mathrm{e/o}$, cf.~\Eq{LambdaReopMain}, and thus $\Lambda^{R}_\mathrm{e/o}$ scales as $(\Pi^R_\mathrm{e/o})^{-1}$, giving the scaling exponent of the $T$ matrix,
\begin{align}
	m=d+z\,.
\end{align}
Combining all scaling relations quoted above,
the exponents $\alpha$ and $\beta$, cf.~Eq. \eqref{eq:exponents}, are found to be
\begin{align}
	\beta=\frac{1}{z+d}=\frac{1}{2+d}
	\,,
	\qquad
	\alpha=\frac{d}{z+d}
	\,.
	\label{eq:alphabetaanomalous}
\end{align}

These can be compared with the standard `Gaussian' exponents, which are reproduced here for the case of small $F_{0}$, for which the term $n=1$ dominates the collision integral, 
\begin{align}
\label{eq:betamuG}
	\mu_\mathrm{G}
	&=d+2n(d-\alpha/\beta)-d-z+2m-\alpha/\beta
	=2m-z-\alpha/\beta
	\,,
\end{align}
since one more $\mathbf{q}_{i}$ integral contributes at each order $n$ of the integral \eqref{app-eq:Cfmain2}, which is due to the fact that in elastic two-to-two scattering away from pure forward scattering (leading to a wave-turbulent cascade with $\kappa=d$) all four momenta must be $q_{i}\gg p_{\Lambda}$ \cite{Chantesana:2018qsb.PhysRevA.99.043620}.
Analogously, the scaling of the $T$-matrix is given by
$m_\mathrm{G}=z$,
such that the  exponents $\alpha$ and $\beta$ for this `Gaussian' non-thermal fixed point \cite{Karl2017b.NJP19.093014,Chantesana:2018qsb.PhysRevA.99.043620,Mikheev:2018adp} result as
\begin{align}
	\beta_\mathrm{G}=\frac{1}{z}=\frac{1}{2}
	\,,
	\qquad
	\alpha_\mathrm{G}=\frac{d}{z}
	\,.
	\label{eq:alphabetagaussian}
\end{align}
%

\subsection{Spatial scaling form}
The exponent $\kappa$ character\bae{is}{iz}ing the scaling function 
$f_\mathrm{s}(\mathbf{p})\sim |\mathbf{p}|^{-\kappa}$ is derived in a similar manner.
Defining the scaling dimension of the scattering integral at a fixed time $t_{0}$ through
$C[f](t_{0},\mathbf{p})=s^{-\mu_\kappa} C[f]\left(t_{0},s \mathbf{p}\right)$,
the time-independent fixed-point equation
\begin{align}
	\label{eq:ScalingFPEq}
	\left(\alpha +\beta\mathbf{p}\cdot\partial_{\mathbf{p}}\right) f_\mathrm{s}(\mathbf{p})
	=t_{0}C[f_\mathrm{s}](\mathbf{p})
	\,
\end{align}
then demands that 
$\kappa=-\mu_\kappa$,
as long as $\kappa\not=d$, which would give a stationary wave-turbulent cascade, cf.~\cite{Chantesana:2018qsb.PhysRevA.99.043620}.
Rescaling every momentum in  the scattering integral \eqref{app-eq:Cfmain2} as $\mathbf{p}\to s\mathbf{p}$ on finds that, the order-$n$ term rescales according to 
\begin{align}
	\label{eq:prelimmukappa2} 
        C^{(n)}[f](t_{0},\mathbf{p};p_{\Lambda})
        &=s^{-2nd-2m_\kappa+d+z+(2n+1)\kappa} C^{(n)}[f]\left(t_{0},s \mathbf{p};sp_{\Lambda}\right)
	\,,
\end{align}
where again, the `free' momentum $\mathbf{k}$ is considered to not contribute to the scaling, and where the scaling dimension of the $T$-matrix at fixed time is $m_\kappa$, i.e.,
$|T^{(n)}(\mathbf{p},\{\mathbf{q}_{i}\})|
	= s^{-m_\kappa} |T^{(n)}(s\mathbf{p},\{s\mathbf{q}_{i}\})|$.	
In order to have the scaling form \eqref{eq:fMK} be homogeneous, one also needs to rescale $p_{\Lambda}$. 
A large number of momenta can take values $q_{i}\lesssim p_{\Lambda}$, and such configurations are expected to dominate the scattering integral because the scaling form takes its largest value there. 
In this momentum regime, the respective integrals do not contribute to scaling in $\mathbf{p}$.
In fact, the integral must be dominated, in each summand, by all but a few momenta being evaluated below $p_{\Lambda}$ because otherwise, the scattering integral would not be a homogeneous function of $\mathbf{p}$. 
This introduces, however, a contribution to the scaling of $C^{(n)}$ which originates from the cutoff scale $p_{\Lambda}$ alone and which distorts the scaling dimension $\mu_{\kappa}$.
This scaling is given by
$C^{(n)}[f](t_{0},\mathbf{p};p_{\Lambda})=s^{-(2n-1)(d-\kappa)} C^{(n)}[f](t_{0},\mathbf{p};sp_{\Lambda})$,
since, in each term, if $\kappa>d$, there are at most $2n-1$ algebraically divergent integrals over functions $f\sim p^{-\kappa}$, which each contribute a leading-order dependence $\sim p_{\Lambda}^{d-\kappa}$ on the cutoff, cf.~\cite{Heinen2022b} for a more details and \cite{Chantesana:2018qsb.PhysRevA.99.043620} for the respective discussion for the case of a $\lambda\varphi^{4}$ model.

Subtracting the degree of divergence $(2n-1)(\kappa-d)$ from the scaling dimension of $C^{(n)}$, one obtains the $n$-independent exponent 
\begin{align}
	\label{eq:mukappa}
	\mu_\kappa=2m_\kappa-z-2\kappa
	\,.
\end{align}

A similar analysis for the $n$th-order contribution to the $T$-matrix leads to the scaling dimension
$m_\kappa=z+d$,
which leads to the final result, cf.~\eqref{appeq:scalrel},
\begin{align}
	\kappa=2d+z=2d+2
	\,.
	\label{eq:kappaanomalous}
\end{align}
Again, this is in contrast with the Gaussian fixed point, where one finds the weaker fall-off with, cf.~Ref.~\cite{Chantesana:2018qsb.PhysRevA.99.043620},
\begin{align}
	\kappa_\mathrm{G}=d+z/2=d+1
	\,.
\end{align}
%

\subsection{Non-locality of transport in sine-Gordon kinetics}
\label{sec:nonlocality-in-kinetics}
Given the kinetics behind scaling as sketched in the previous two subsections, we finally return to the discussion of the non-locality of transport introduced in \Sect{nonlocality} above.
The scaling relation $\alpha=d\beta$ ensures that all terms in the sums contributing to the scattering integral \eqref{app-eq:Cfmain2} and the loop functions \eq{PiReQuasiP}f.~entering the $T$-matrix show the same scaling in space and time. 
In contrast, as discussed in the last subsection, homogeneity of the scattering integral at a fixed time $t_{0}$, \eq{prelimmukappa2} is ensured by the integral and loop functions being dominated by configurations where $2n-1$ of the momenta $\mathbf{q}_{i}$ are evaluated at momenta $\lesssim p_{\Lambda}$, where the distribution function $f_{\mathbf{q}_{i}}$ reaches its maximum, plateau value, such that the respective integrals do not contribute to the rescaling of $C[f]$ when rescaling its argument $\mathbf{p}\to s\mathbf{p}$.
The latter implies that the homogeneity index of $C^{(n)}[f](t_{0},\mathbf{p})$ as well as of the loop functions is reduced by $(2n-1)(d-\kappa)$, which, as discussed above, leads to the scaling exponent $\kappa=2d+2$. 
Moreover, this means that of the in total $2n+2$ momenta in the $n$th summand, only $3$ are of order $\gg p_{\Lambda}$.

Hence, if the scattering integral is dominated by terms of the order $n\approx F_{0}\gg1$, in each elastic collision process that redistributes the momentum occupations, many of the momenta are within the plateau regime while only $3$ contribute to the transport at momenta $\gg p_{\Lambda}$ including those where the energy is deposited within the UV.
The many low, $\mathcal{O}(p_{\Lambda})$ momenta are expected to contribute to the strongly peaked current depicted in Fig.~\ref{fig:SourcesJQ}, while the remaining three modes give rise to the particularly steep power law $\kappa=2d+2$ governing the scaling function and thus the much steeper fall-off of the current for $p\gg p_{\Lambda}$.
	

\end{appendix}


%

\end{document}